\documentclass{osa-article}

\journal{osajournal}

\articletype{Research Article}
\usepackage[utf8]{inputenc}
\usepackage[british]{babel}
\usepackage{amsmath,amsfonts}
\usepackage{cancel}
\usepackage{bm}
\usepackage{comment}
\usepackage[utf8]{inputenc}
\usepackage{xcolor}
\usepackage{graphicx}
\usepackage{mathtools}
\usepackage{float}
\usepackage{enumitem}
\usepackage{soul}
\usepackage{graphics, setspace}
\usepackage{hyperref}
\usepackage{pgf,tikz}
\usetikzlibrary{arrows}
\usetikzlibrary{shapes}
\usetikzlibrary{intersections}
\usetikzlibrary{calc}
\usetikzlibrary{math}
\usetikzlibrary{decorations.markings}
\newcommand\numberthis{\addtocounter{equation}{1}\tag{\theequation}}

\definecolor{tuewarmred}{rgb}{0.969,0.192,0.192}
\definecolor{tueprocesscyan}{rgb}{0.000,0.635,0.871}
\definecolor{tuecyan}{rgb}{0.000,0.635,0.871}
\definecolor{tuered}{rgb}{0.839,0.000,0.290}
\definecolor{tueblue}{rgb}{0.000,0.400,0.800}
\definecolor{tuedarkblue}{rgb}{0.063,0.063,0.451}
\definecolor{tueorange}{rgb}{1.000,0.604,0.000}
\definecolor{tueyellow}{rgb}{1.000,0.867,0.000}
\definecolor{tuelightgreen}{rgb}{0.518,0.824,0.000}
\definecolor{tuegreen}{rgb}{0.000,0.675,0.510}

\begin{document}

\title{Alternative computation of the Seidel aberration coefficients using the Lie algebraic method}

\author{A. Barion,\authormark{1,*} M. J. H. Anthonissen,\authormark{1}
        J. H. M. ten Thije Boonkkamp,\authormark{1} and
        W. L. IJzerman\authormark{1,2}}

\address{\authormark{1}Eindhoven University of Technology, PO Box 513, 5600 MB Eindhoven, The Netherlands\\
\authormark{2}Signify, High Tech Campus 7, 5656 AE Eindhoven, The Netherlands}

\email{\authormark{*}a.barion@tue.nl} 



\begin{abstract}
We give a brief introduction to Hamiltonian optics and Lie algebraic methods. We use these methods to describe the operators governing light propagation, refraction and reflection in phase space. The method offers a systematic way to find aberration coefficients of any order for arbitrary rotationally symmetric optical systems. The coefficients from the Lie method are linked to the Seidel aberration coefficients. Furthermore, the property of summing individual surface contributions is preserved by the Lie algebraic theory. Three examples are given to validate the proposed methodology with good results.
\end{abstract}

\section{Introduction}
	
	The calculation of Seidel aberrations of rotationally symmetric optical systems is well known in the optical design literature \cite{Welford1986,Braat2019}. Seidel sums provide a clear insight into the contribution of each optical surface of the system at hand to the third order transverse ray aberrations at the image plane. Closed form expressions for higher order aberration coefficients are increasingly more complicated \cite{Andersen, Bociort}. While different works in literature only work with the expansion of the ray tracing equations \cite{Andersen, Bociort}, the approach in this paper will show how to approximate the mapping of the optical system such that, when applied to the object plane variables, it produces the aberration expansion up to the desired order in terms of initial position and direction. This will lead to a mathematical framework that allows us to systematically derive the aberration coefficients in closed form and that can be applied to more general optical systems.
	
	In this paper we aim to deliver a concise introduction to the Lie algebraic method, introduced and mainly developed by Alex Dragt and Kurt Bernardo Wolf \cite{DragtFinn,DragtFoundations86,Wolf2004,Dragt82,SaadWolf1986}, applied to geometrical optics and to link this formulation to the widely known Seidel coefficients. In \cite{Dragt82} a first step towards the classification of optical elements using the Lie method has been made. In our current work we introduce the treatment of a fundamental optical map in the framework of the Lie method. The corresponding fundamental optical element is composed of propagation from an intermediate object plane, refraction or reflection by an interface and subsequent propagation to an intermediate image plane. 
	By considering rotationally symmetric optical systems as a concatenation of fundamental optical elements, we will reduce the analysis of the complete system to the determination only the fundamental map, thus significantly simplifying the treatment of arbitrary rotationally symmetric systems. 
	
	The Lie approach provides a mathematical theory which enables us to systematically compute aberration coefficients of a rotationally symmetric system and link them to the well-known Seidel coefficients. It will be shown how the property of summing up individual surface contributions to the Seidel coefficients, as in traditional optical design theory \cite{Welford1986,Braat2019}, is a direct consequence of the presented mathematical framework.
	
	In Section \ref{sec::ClassicForm} and Section \ref{sec::LieTools}, a brief description of geometrical optics and the Lie algebraic approach will be given. In Section \ref{sec::LieForm} the reader will find the methodology necessary to formulate propagation, refraction and reflection within the context of the Lie algebraic method. The description of the fundamental element can be found in Section \ref{sec::refrElandSys}. An optical system can then be viewed as the concatenation of multiple optical elements as described in Section \ref{sec::OptSys}. The link between the Lie framework and the Seidel coefficients is presented in Section \ref{sec::LieToSeid}. Three examples to validate the presented method will be shown in Section \ref{sec::Examples} with very satisfactory  results. Section \ref{sec::Conclusions} offers concluding observations and an outlook to future applications to off-axis systems.
	
	\section{Classical Formulation of Geometrical Optics in Phase Space}\label{sec::ClassicForm}
	In this section, we will summarize the classical theory of geometrical optics describing light propagation, refraction and reflection. These results are what we aim to reproduce in Section \ref{sec::LieForm} and Section \ref{sec::refrElandSys} with the use of Lie algebraic tools.
	
	First, we describe free propagation of a light ray. Consider a general ray in three-dimensional space with coordinates $(q_1,q_2,z)$ and let $s$ represent the arc length along said ray. We can then describe the ray path using $z$ as the propagation parameter for $q_1(z)$ and $q_2(z)$.
	Let $\bm{P}^\mathrm{i}$ and $\bm{P}^\mathrm{f}$ be the initial and final point, respectively, of a ray path $\mathcal{C}$, then its optical path length is defined as 
	\begin{equation}
		\label{eq::opticalPath}
		\text{OPL}=\int_{\mathcal{C}}n(\bm{q},z)\,\mathrm{d}s=\int_{z^\mathrm{i}}^{z^\mathrm{f}}n(\bm{q},z)\sqrt{1+\vert\dot{\bm{q}}\vert^2}\,\mathrm{d}z,
	\end{equation}
	where $\bm{q}=(q_1,q_2)$, $\bm{P}^\mathrm{i}=\bm{P}^\mathrm{i}(\bm{q}(z^\mathrm{i}),z^\mathrm{i}),\,\bm{P}^\mathrm{f}=\bm{P}^\mathrm{f}(\bm{q}(z^\mathrm{f}),z^\mathrm{f})$ and $n(\bm{q},z)$ is the refractive index of the medium through which the ray propagates. Fermat's Principle tells us that the ray path is a stationary point of Eq.~\eqref{eq::opticalPath} \cite{Luneburg}. As such, the ray path satisfies the Euler-Lagrange equations \cite{Luneburg}
	\begin{subequations}
		\label{eq::Lagrangian}
		\begin{equation}
			\frac{\partial\mathcal{L}}{\partial \bm{q}}-\frac{\mathrm{d}}{\mathrm{d}z}\frac{\partial\mathcal{L}}{\partial \dot{\bm{q}}}=\bm{0},
		\end{equation}
	with the Lagrangian $\mathcal{L}$ given by
		\begin{equation}
			\mathcal{L}(\bm{q},\dot{\bm{q}},z)=n(\bm{q},z)\sqrt{1+\vert\dot{\bm{q}}\vert^2}.
		\end{equation}
	\end{subequations} 
 	The canonical conjugates of $\dot{\bm{q}}$, also called momenta, read
	\begin{equation}
		\label{eq::momenta}
		\bm{p}=\frac{\partial\mathcal{L}}{\partial\dot{\bm{q}}}=n(\bm{q},z)\frac{\dot{\bm{q}}}{\sqrt{1+\vert\dot{\bm{q}}\vert^2}}.
	\end{equation}
	The conjugates $\bm{p}=(p_1, p_2)$ are the projections of the unit direction vector of the ray onto a plane $z=\mathrm{const.}$, e.g., $z=0$, times the refractive index. 
	The Hamiltonian $H$ is defined in terms of $\mathcal{L}$ by the relation:
	\begin{equation}
		H(\bm{q},\bm{p},z)=\bm{p}\boldsymbol{\cdot}\dot{\bm{q}}-\mathcal{L}(\bm{q},\dot{\bm{q}},z),\label{eq::ham}
	\end{equation}
	where $\dot{\bm{q}}$ is expressed in terms of $\bm{q},\bm{p},z$ through a Legendre transformation \cite{Luneburg}.
	
	Combining the Lagrangian system \eqref{eq::Lagrangian} with the definition of $\bm{p}$ and $H$ in Eq.~\eqref{eq::momenta} and Eq.~\eqref{eq::ham}, respectively, we obtain
	
	\begin{subequations}
		\label{eq::HamSys}
		\begin{equation}
			\label{eq::HamEq}
			\dot{\bm{q}}=\frac{\partial H}{\partial\bm{p}},\quad\dot{\bm{p}}=-\frac{\partial H}{\partial\bm{q}},
		\end{equation}
		where the Hamiltonian $H$ is given by
		\begin{equation}
			H(\bm{q},\bm{p},z)=-\sqrt{n(\bm{q},z)^2-\vert\bm{p}\vert^2}.\label{eq::ham2}
		\end{equation}
	\end{subequations}
	A ray is completely defined by its position and direction coordinates along the $z$-axis. Hence, its initial values $(\bm{q}^\mathrm{i},\bm{p}^\mathrm{i})$ at the plane $z=z^\mathrm{i}$ are related to its final coordinates at the plane $z=z^\mathrm{f}$ by following the trajectory $\big(\bm{q}(z),\bm{p}(z),z\big)$ governed by the Hamiltonian system \eqref{eq::HamSys}. 
	From now on we will refer to $\bm{q},\bm{p}$ as phase space variables. Let $p_z$ be the $z$-component of the unit direction vector along the ray, multiplied by the refractive index $n$. In an optical medium it holds true that $n(\bm{q}(z),z)=\vert\big(\bm{p}(z),p_z(z)\big)\vert$. As such, $p_z$ can be defined as
	\begin{equation}
		p_z=\sqrt{n(\bm{q},z)^2-\vert\bm{p}\vert^2},
	\end{equation}
	and by substitution into Eq.~\eqref{eq::ham2} we get $H=-p_z$. We only consider the positive $z$-direction, since we assume rays to be travelling only forward. Furthermore, note that if the Hamiltonian is independent of $z$, then it is constant and thus also $p_z$ is.
	
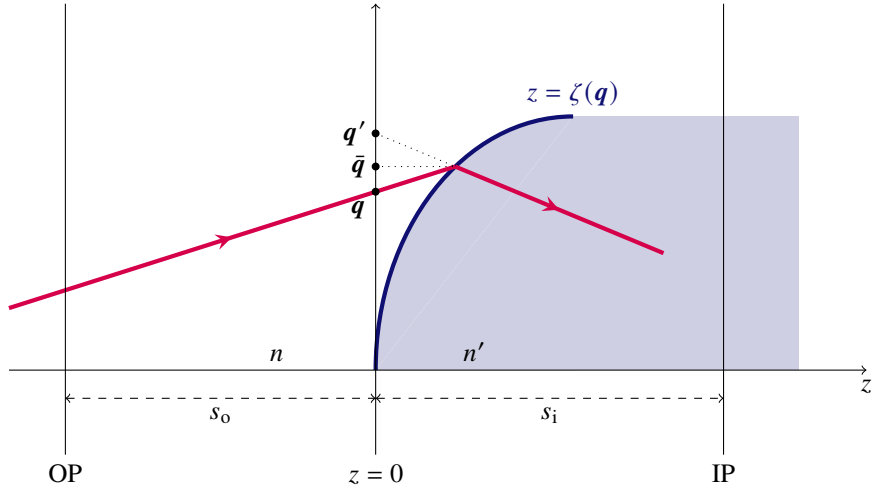
\begin{figure}
		\centering
		\def\scale{1.5}
		\begin{tikzpicture}[scale=\scale]
			
			\tikzmath{
				\z = 0; 
				\d = 4; 
				%
				\la = 1.75;
				\lb = 2.25;
				\lMx = \z; 
				\lMy = 0;
			}
			
			\fill [tuedarkblue!20] (\z,\lb) arc (90:180:{\la} and {\lb});
			\fill [tuedarkblue!20] (\z,\lb) -- (\z+\d/2,\lb) -- (\z+\d/2,-0) -- (\z,-0) -- cycle;
			\fill [tuedarkblue!20] (\z-\la,0) -- (\z,\lb) -- (\z,-0) -- cycle;
			
			\draw [ultra thick,tuedarkblue,name path=leftsurface] (\z,\lb)  node [above] {$z = \zeta(\bm{q})$} arc (90:180:{\la} and {\lb});
			
			
			\def\arrowpos{0.5}; 
			\begin{scope}[decoration={markings,mark=at position \arrowpos with {\arrow{stealth}}}]
				\foreach \y in { 0.55 }{
					\path [name path=rayin] (-5,\y) -- (\z,0.95*\lb);
					\path [name intersections={of=rayin and leftsurface, by=A}]; 
					\draw [tuered, ultra thick, postaction={decorate}] (-5,\y) -- (A);  
					
					\newdimen\XCoord
					\newdimen\YCoord
					\pgfgetlastxy{\XCoord}{\YCoord}; 
					
					\tikzmath{
						\Ax = \XCoord*1pt/(1cm*\scale) - \lMx;
						\Ay = \YCoord*1pt/(1cm*\scale) - \lMy;
						\nx = \lb^2 * \Ax;
						\ny = \la^2 * \Ay; 
						\vectornorm = sqrt((\nx)^2 + (\ny)^2);
						\nx = \nx / \vectornorm;
						\ny = \ny / \vectornorm;
						\taux = -\ny;
						\tauy = \nx;
						\sx = 1;
						\sy = 0;
						\e = 1 / 1.8; 
						\c = -\nx*\sx-\ny*\sy;
						\b = \e*\c - sqrt(1-(\e)^2*(1-(\c)^2));
						\tx = \e*\sx + \b*\nx;
						\ty = \e*\sy + \b*\ny;
						\m = \z+\d/2;
					}
					
					
					\draw [tuered, ultra thick, postaction={decorate}] (A) -- ++({\m*\tx},{\m*\ty});
					
					\draw [dotted] (\z-\la,\Ay) -- (A);
					\fill (\z-\la,\Ay) circle(1pt) node [left] {$\bar{\bm{q}}$};
					
					\path [name path=verticalaxis] (\z-\la,-0.75) -- (\z-\la,3);
					\path [name intersections={of=rayin and verticalaxis, by=Q}];			
					\fill (Q) circle(1pt) node [below left] {$\bm{q}$};
					
					\path [name path=refractedray] (A) -- ++({-\m*\tx},{-\m*\ty});
					\path [name intersections={of=refractedray and verticalaxis, by=Qprime}];		
					\draw [dotted] (Qprime) -- (A);	
					\fill (Qprime) circle(1pt) node [left] {$\bm{q}'$};
				}
				
				\draw [->] (\z-\la,-0.75) node [below] {$z=0$} -- ++(0,4);
				\draw [->] (-5,0) -- (\z+1.3*\d/2,0) node [below] {$z$};
				\draw [-] (-4.5,-0.75) node [below] {OP} -- ++(0,4);
				\draw [-] (\z+\d/3,-0.75) node [below] {IP} -- ++(0,4);
				
				\draw [<->, dashed] (-4.5,-0.25) -- (\z-\la,-0.25) node [midway, below] {$s_\mathrm{o}$};
				\draw [<->, dashed] (\z-\la,-0.25) -- (\z+\d/3,-0.25) node [midway, below] {$s_\mathrm{i}$};

				\draw (\z-\la/2,0) node [above] {$n'$};
				\draw (\z-3*\la/2,0) node [above] {$n$};

			\end{scope}
		\end{tikzpicture}
		\caption{Refraction at the surface $z=\zeta(\bm{q})$ where $n'>n$. Position coordinates are projected along the ray onto the standard screen $z=0$. Intermediate object and image planes are displayed.}
		\label{fig::rootTransf}
	\end{figure}

	Next, we continue this section with the refraction of rays, where we will consider the refractive index to be constant within each respective medium. To this purpose, we introduce two quantities that are conserved at refraction: the position coordinates $\bar{\bm{q}}$ of the point of impact of the ray with the refracting surface and the tangential momentum $\bar{\bm{p}}$ at this point. To this purpose we denote with the unprimed values $(\bm{q},\bm{p})$ the phase space coordinates of the incoming ray, with $(\bar{\bm{q}},\bar{\bm{p}})$ the conserved ray quantities at the refracting interface and with $(\bm{q}',\bm{p}')$ the phase space coordinates of the outgoing ray. The position coordinates are measured by (back-)propagating the ray onto the so-called standard screen defined by $z=0$, see Figure \ref{fig::rootTransf}. Let $z=\zeta(\bm{q})$ be the refracting surface, then the conservation of the point of impact is described through ray propagation:
	\begin{equation}
		\label{eq::posConservation}
		\bm{q}+\zeta(\bar{\bm{q}})\frac{\bm{p}}{p_z}=\bar{\bm{q}}=\bm{q}'+\zeta(\bar{\bm{q}})\frac{\bm{p}'}{p_z'}.
	\end{equation}
	This is the solution of the Hamiltonian system \eqref{eq::HamSys} in a medium of constant refractive index subject to $\bm{q}(0)=\bm{q}, \bm{p}(0)=\bm{p}$.
	Note that the standard screen can be placed at any distance from the optical surface. However, the assumption that the standard screen and the surface both intersect at $\bm{q}=\bm{0}$ simplifies the expressions in Eq.~\eqref{eq::posConservation}.

	Conservation of tangential momentum can be described through the following cross product between the surface normal, to be taken opposite to the incoming rays, and the ray direction coordinates. Let $\nabla\zeta(\bar{\bm{q}})$ be the gradient of our refracting surface at the point of impact, then
	\begin{equation}
		\binom{\nabla\zeta(\bar{\bm{q}})}{-1}\boldsymbol{\times}\binom{\bm{p}}{p_z}=\binom{\nabla\zeta(\bar{\bm{q}})}{-1}\boldsymbol{\times}\binom{\bm{p}'}{p_z'}.
	\end{equation}
	The first two components give 
	\begin{equation}
		\label{eq::momConservation}
		\bm{p}+\nabla\zeta(\bar{\bm{q}})p_z=\bar{\bm{p}}=\bm{p}'+\nabla\zeta(\bar{\bm{q}})p_z'.
	\end{equation}
	With the results in Eq.~\eqref{eq::posConservation} and Eq.~\eqref{eq::momConservation} we can define the root transformations $\mathcal{R}_{n;\zeta}$ that acts upon the incident ray coordinates and gives us the conserved quantities $(\bar{\bm{q}},\bar{\bm{p}})$ at the interface \cite{Wolf2004,SaadWolf1986}:
	\begin{subequations}
	\label{eq::rootTransf}
	    \begin{equation}
	    \label{eq::rootTransfQ}
			\mathcal{R}_{n;\zeta}\bm{q}=\bar{\bm{q}}=\bm{q}+\zeta(\bar{\bm{q}})\frac{\bm{p}}{\sqrt{n^2-\vert\bm{p}\vert^2}},
		\end{equation}
		\begin{equation}
		\label{eq::rootTransfP}
			\mathcal{R}_{n;\zeta}\bm{p}=\bar{\bm{p}}=\bm{p}+\nabla\zeta(\bar{\bm{q}})\sqrt{n^2-\vert\bm{p}\vert^2}.    
		\end{equation}
	\end{subequations}
	Note that these definitions are implicit, because the expression for $\bar{\bm{q}}$ is dependent on $\bar{\bm{q}}$ itself. Once the position coordinate of the point of impact is known, then the expression for $\bar{\bm{p}}$ can be computed explicitly. The root transformation $\mathcal{R}_{n';\zeta}$ is defined analogously to Eqs.~\eqref{eq::rootTransf}, but has the refracted phase space coordinates $(\bm{q}',\bm{p}')$ and the refractive index $n'$ as arguments, see Eq.~\eqref{eq::posConservation} and Eq.~\eqref{eq::momConservation}. In this formulation we can summarize the conservation at the point of impact as 
	\begin{equation}
		\mathcal{R}_{n;\zeta}\binom{\bm{q}}{\bm{p}}=\binom{\bar{\bm{q}}}{\bar{\bm{p}}}=\mathcal{R}_{n';\zeta}\binom{\bm{q}'}{\bm{p}'}.
	\end{equation}
	The inverse transformation of $\mathcal{R}_{n';\zeta}$ is derived from Eq.~\eqref{eq::posConservation} and Eq.~\eqref{eq::momConservation} by bringing to the other side the necessary terms and reads:
	\begin{subequations}
	    \label{eq::invRootTransf}
		\begin{equation}
    		\mathcal{R}^{-1}_{n';\zeta}\bar{\bm{q}}=\bm{q}'=\bar{\bm{q}}-\zeta(\bar{\bm{q}})\frac{\bm{p}'}{\sqrt{n'^2-\vert \bm{p}'\vert^2}},
    	\end{equation}
    	\begin{equation}
    		\mathcal{R}^{-1}_{n';\zeta}\bar{\bm{p}}=\bm{p}'=\bar{\bm{p}}-\nabla\zeta(\bar{\bm{q}})\sqrt{n'^2-\vert \bm{p}'\vert^2}.
	    \end{equation}
	\end{subequations}
	This is again defined implicitly, since the expression for $\bm{p}'$ is dependent on $\bm{p}'$ itself. We are interested in the refraction mapping $S_{n,n';\zeta}$ between the incoming and outgoing phase space coordinates after refraction at the standard screen. This is achieved by combining the root transformations in Eqs.~\eqref{eq::rootTransf} and their inverses in Eqs.~\eqref{eq::invRootTransf} into \cite{Atzema,DragtFoundations86,Wolf2004}
	\begin{equation}
	\label{eq::refrOperator}
		S_{n,n';\zeta}=\mathcal{R}^{-1}_{n';\zeta}\circ\mathcal{R}_{n;\zeta}.
	\end{equation} 
	
	At this point we assume that the surface intersects the standard screen at the origin, i.e., $\zeta(\bm{0})=0$, and that the surface is tangential to the standard screen at the origin, i.e., $\nabla\zeta(\bm{0})=\bm{0}$. The first assumption is without any loss of generality since the root transformations can be reformulated according to the position of the standard screen, which can therefore be moved freely along the $z$-axis. The second assumption limits the possible surface representations to functions with no first order terms in their Taylor expansion at $\bm{q}=\bm{0}$. The motivation behind these assumptions will become apparent in the following sections, but the first immediate consequence is that the origin of phase space is mapped onto itself by $S_{n,n';\zeta}$, i.e., if $\bm{q}=\bm{0}=\bm{p}$ then $\bm{q}'=\bm{0}=\bm{p}'$.
	
	The implicit form of the root transformations makes it necessary to compute them recursively up to the desired order of accuracy. We start with a fixed point iteration using $\bar{\bm{q}}=\bm{q}$ as our first guess. Solving Eq.~\eqref{eq::rootTransfQ} iteratively allows us to find an expansion for $\bar{\bm{q}}$ up to the desired order. Substituting this result in Eq.~\eqref{eq::rootTransfP} gives us an expansion for $\bar{\bm{p}}$. We then proceed similarly to derive expansions of $\bm{q}',\bm{p}'$ in terms of $\bar{\bm{q}},\bar{\bm{p}}$ using Eqs.~\eqref{eq::invRootTransf}. Combining these results yields an expansion of the outgoing ray coordinates in terms of the incoming ones up to the desired order of accuracy. A rotationally symmetric refracting surface can be described by
	\begin{equation}
		\label{eq::rotSurf}
		z=\zeta(\bm{q})=\beta\vert\bm{q}\vert^2+\delta(\vert\bm{q}\vert^2)^2+\cdots.
	\end{equation}
	If we consider the surface description in Eq.~\eqref{eq::rotSurf}, then the third order expansion of $S_{n,n';\zeta}$ in Eq.~\eqref{eq::refrOperator} applied to the phase space variables of an incoming ray reads:
	\begin{subequations}
		\label{eq::primed}
		\begin{align}
			\bm{q}'=&\,\bm{q}+\beta\left(\frac{1}{n}-\frac{1}{n'}\right)\vert\bm{q}\vert^2\bm{p}+2\beta^2\left(1-\frac{n}{n'}\right)\vert\bm{q}\vert^2\bm{q}\label{eq::qPrime},\\
			\bm{p}'=&\,\bm{p}+2\beta(n-n')\bm{q}-\beta\left(\frac{1}{n}-\frac{1}{n'}\right)\vert\bm{p}\vert^2\bm{q}+2\beta^2\left(1-\frac{n'}{n}\right)\vert\bm{q}\vert^2\bm{p}\nonumber\\
			&-4\beta^2\left(1-\frac{n}{n'}\right)(\bm{p}\cdot \bm{q})\bm{q}	+4(n-n')\left(\delta+\beta^3\frac{n-n'}{n'}\right)\vert\bm{q}\vert^2\bm{q},\label{eq::pPrime}
		\end{align}
	\end{subequations}
	and is in agreement with \cite{SaadWolf1986,DragtFoundations86}. The linear part of Eqs.~\eqref{eq::primed} describes refraction within the limits of Gaussian optics \cite{Hecht}. We will refer to the linear components of propagation and refraction as the Gaussian part of said mappings.
	
	In the case of a reflecting surface similar calculations can be performed. The corresponding reflection relations read \cite{Wolf2004}:
	\begin{equation}
		\begin{gathered}
			\bm{q}+\zeta(\bar{\bm{q}})\frac{\bm{p}}{\sqrt{n^2-\vert \bm{p}\vert^2}}=\bar{\bm{q}}=\bm{q}'-\zeta(\bar{\bm{q}})\frac{\bm{p}'}{\sqrt{n^2-\vert \bm{p}'\vert^2}},\\
			\bm{p}+\nabla\zeta(\bar{\bm{q}})\sqrt{n^2-\vert \bm{p}\vert^2}=\bar{\bm{p}}=\bm{p}'-\nabla\zeta(\bar{\bm{q}})\sqrt{n^2-\vert \bm{p}'\vert^2}.
		\end{gathered} 
		\label{eq::reflRoots}
	\end{equation}
	We notice that Eqs.~\eqref{eq::reflRoots} consider a back-propagation of the reflected ray compared to Eqs.~\eqref{eq::rootTransf}. Following the steps presented in the case of refraction, similar expressions to Eqs.~\eqref{eq::primed} can be calculated. A way to promptly derive the reflection expansion from Eqs.~\eqref{eq::primed} is to substitute $n'=-n$ in Eqs.~\eqref{eq::primed} \cite{Atzema}, but we encourage the reader to perform the calculations themselves to get better acquainted with the procedure. We will focus on the treatment of refractive optics, but each step made using Eqs.~\eqref{eq::rootTransf} can be applied analogously to Eq.~\eqref{eq::reflRoots}.

	\section{Lie Algebraic Tools}\label{sec::LieTools}
	This section serves as a brief introduction to the theory of Lie algebraic tools needed to describe light propagation and refraction introduced in Section \ref{sec::ClassicForm}. We first define the Poisson bracket and the Lie algebra structure it generates for functions on phase space. Next, the concepts of Lie operators and Lie transformations are explained. Important results connecting Lie transformations and symplectic mappings are presented together with some essential tools for the treatment of Lie transformation products. From this section onward all functions on phase space are assumed to be sufficiently smooth for any differentiation they might be subjected to.
	
	The Poisson bracket is an operator $[\cdot,\cdot]$, which maps any pair of functions $f,g$ in $(\bm{q},\bm{p})$ to a single function of $(\bm{q},\bm{p})$, denoted by $[f,g]$:
	\begin{equation}
	\label{eq::poissBracket1}
		[ f,g]:=\sum_{i=1}^2 \left(\frac{\partial f}{\partial q_i}\,\frac{\partial g}{\partial p_i}-\frac{\partial f}{\partial p_i}\,\frac{\partial g}{\partial q_i}\right)=\frac{\partial f}{\partial \bm{q}}\boldsymbol{\cdot}\frac{\partial g}{\partial \bm{p}}-\frac{\partial f}{\partial \bm{p}}\boldsymbol{\cdot}\frac{\partial g}{\partial \bm{q}}.
	\end{equation}
	The above definition of the Poisson bracket implies that $[q_i,p_j]=\delta_{ij}$. The Poisson bracket has the following properties:
	\begin{subequations}
		\begin{align}
			&[\alpha f +\beta g, h]=\alpha[f,h]+\beta[g,h]&\text{(linear in first component)},\\
			&[f,\alpha g +\beta h]=\alpha[f,g]+\beta[f,h]&\text{(linear in second component)},\\
			&[f,f]=0&\text{(alternating)},\label{eq::firstLie}\\
			&[f,[g,h]]+[g,[h,f]]+[h,[f,g]]=0&\text{(Jacobi identity)}\label{eq::secondLie}.
		\end{align}
	\end{subequations}
	As a consequence of Eq.~\eqref{eq::firstLie}, the Poisson bracket is also antisymmetric, i.e., $[f,g]=-[g,f]$. Thus, the Poisson bracket turns the space of functions defined on phase space into a Lie algebra \cite{Fuchs}.

	The transformation to new phase space variables $(\bm{q},\bm{p})\mapsto (\bm{Q}(\bm{q},\bm{p})$, $\bm{P}(\bm{q},\bm{p}))$ is said to be a canonical transformation, if it satisfies \cite{Wolf2004,DragtFinn}:
	\begin{align}
		[Q_i,Q_j]&=[q_i,q_j]=0,\nonumber\\
		[P_i,P_j]&=[p_i,p_j]=0,\label{eq::canonicalTransformation}\\
		[Q_i,P_j]&=[q_i,p_j]=\delta_{ij}.\nonumber
	\end{align}
	Consequently, the value of Poisson brackets is preserved under canonical transformations \cite{Quillen}. We now collect the two sets of variables $q_i,p_i$ and $Q_i,P_i$, into a single set of four variables $w_1,\ldots,w_{4}$ and $W_1,\ldots,W_4$, respectively, where
	\begin{equation}
	\label{eq::wVars}
		\bm{w}=(q_1,q_2,p_1,p_2)^T,\quad\bm{W}=(Q_1,Q_2,P_1,P_2)^T.
	\end{equation}
	Using the notation of Eq.~\eqref{eq::wVars}, the rules in Eqs.~\eqref{eq::canonicalTransformation} can be summarized by
	\begin{equation}
		[w_i,w_j]=J_{ij},\text{ where } J=\begin{pmatrix}
			O_{\scriptscriptstyle 2\times2}&I_{\scriptscriptstyle 2\times2}\\
			-I_{\scriptscriptstyle 2\times2} & O_{\scriptscriptstyle 2\times2}
		\end{pmatrix}.
		\label{eq::defJ}
	\end{equation}
	The Poisson bracket operation \eqref{eq::poissBracket1} in terms of $\bm{w}$ can be represented as
	\begin{equation}
	\label{eq::poissonBracket2}
		[f,g]=\sum_{k,l=1}^4\frac{\partial f}{\partial w_k}\,J_{kl}\,\frac{\partial g}{\partial w_l}=\left(\frac{\partial f}{\partial \bm{w}}\right)^TJ\,\frac{\partial g}{\partial \bm{w}}.
	\end{equation}
	Let us consider a canonical transformation $\bm{w}\mapsto\bm{W}(\bm{w})$, which therefore satisfies Eqs.~\eqref{eq::canonicalTransformation}. Combining the relations in Eqs.~\eqref{eq::defJ} with the just introduced representation of the Poisson bracket in Eq.~\eqref{eq::poissonBracket2} yields
	\begin{equation}
		J_{ij}=[W_i,W_j]=\sum_{k,l=1}^4\frac{\partial W_i}{\partial w_k}\,J_{kl}\,\frac{\partial W_j}{\partial w_l}.
		\label{eq::symplecticCond}
	\end{equation}
	Let $M$ be the Jacobian matrix of the transformation from $\bm{w}$ to $\bm{W}$, i.e.,
	\begin{equation}
		M_{ik}(\bm{w})=\frac{\partial W_i}{\partial w_k},
		\label{eq::mapJacobian}
	\end{equation}
	then according to Eq.~\eqref{eq::symplecticCond} the following must hold
	\begin{equation}
	\label{eq::symplecticM}
		MJM^T=J.
	\end{equation}
	The relation in Eq.~\eqref{eq::symplecticM} means that the matrix $M$ belongs to the symplectic group in four dimensions \cite{DragtFinn}. Accordingly, the necessary and sufficient condition for a transformation to be canonical is for its Jacobian matrix to be symplectic  and a canonical transformation is therefore often called a symplectic map.
	
	Let $f$ and $g$ be two functions on phase space. We define the linear Lie operator $[f,\cdot\,]$ associated with $f$ acting on $g$ as follows:
	\begin{equation*}
		[f,\cdot\,]g=[f,g].
	\end{equation*}
	If the argument of $[f,\cdot\,]$ is a vector-valued function $\bm{g}$, then it acts component-wise on $\bm{g}$. We now introduce the main Lie algebraic tool used in this work: the Lie transformation. The Lie transformation $\exp([f,\cdot\,])$ associated with $f$ and generated by $[f,\cdot\,]$ is defined as:
	\begin{equation}
		\exp([f,\cdot\,])=\sum_{k=0}^\infty \frac{[f,\cdot\,]^k}{k!}.
		\label{eq::LieTransformation}
	\end{equation}
	The powers in Eq.~\eqref{eq::LieTransformation} follow the recursive definition
	\begin{gather*}
		[f,\cdot\,]^0=I,\\
		[f,\cdot\,]^k=[f,[f,\cdot\,]^{k-1}],\quad k= 1,2,\dots.
	\end{gather*} 
	It can be proven that the Lie transformation associated with a function $f(\bm{q},\bm{p})$ on phase space has the following properties for any differentiable functions $g,h$ on phase space \cite{Wolf2004}:
	\begin{subequations}
	    \begin{align}
	        &\exp([f,\cdot\,])[g,h]=[\exp([f,\cdot\,])g,\exp([f,\cdot\,])h],&\text{(Conservation of Poisson bracket)}\label{eq::poissonCon}\\
	        &\exp([f,\cdot\,])g(\bm{q},\bm{p})=g(\exp([f,\cdot\,])\bm{q},\exp([f,\cdot\,])\bm{p}).&\text{(Action on function arguments)}\label{eq::jumpingIntoArgs}
	    \end{align}
	\end{subequations}

	The motivation for introducing Lie transformations becomes clear with the following two theorems by Dragt and Finn \cite{DragtFinn} which are pivotal for the discussion to follow. They will allow us to establish a direct link between symplectic maps and their representation as Lie transformations. The first theorem \cite[Theorem 1]{DragtFinn} gives us a result regarding maps derived from Lie transformations for a given $f$ of the form: 
	\begin{equation}
	\label{eq::symplecticLieTransf}
		W_i=\exp([f,\cdot\,])w_i.
	\end{equation}
	Provided the series in Eq.~\eqref{eq::symplecticLieTransf} converges, the resulting map is symplectic \cite{DragtFinn}. In fact, using property \eqref{eq::poissonCon} of Lie transformations and their definition in Eq.~\eqref{eq::LieTransformation} we have
    \begin{equation}
    \label{eq::symplecticLieProof}
    \begin{aligned}
        [W_i,W_j]&=[\exp([f,\cdot\,])w_i,\exp([f,\cdot\,])w_j]\\
        &=\exp([f,\cdot\,])[w_i,w_j]\\
        &=\exp([f,\cdot\,])J_{ij}\\
        &=J_{ij},
    \end{aligned}
    \end{equation}
	since $[f,\cdot\,]^kJ_{ij}=0$ for any $k\neq0$. The result in Eqs.~\eqref{eq::symplecticLieProof} states that maps of the form \eqref{eq::symplecticLieTransf} are symplectic. Conversely, the second theorem \cite[Theorem 2]{DragtFinn} gives us a sufficient condition for representing symplectic maps in terms of Lie transformations. Suppose that each component of a symplectic map $\bm{w}\mapsto\bm{W}(\bm{w})$ allows for a representation as
	\begin{equation}
		\label{eq::anSym}
		W_i=\sum_{\vert\sigma\vert>0} a_i(\sigma) w^\sigma,
	\end{equation}
	where $\sigma$ denotes a 4-tuple of exponents in $\mathbb{N}$ and
	\begin{equation}
		\label{eq::vectorNorm}
		\vert\sigma\vert=\sum_{k=1}^{4}\sigma_k,\quad w^\sigma=w_1^{\sigma_1}w_2^{\sigma_2}w_{3}^{\sigma_{3}} w_{4}^{\sigma_{4}}.
	\end{equation}
	Then, $W_i$ can be represented as an infinite product of Lie transformations:
	\begin{equation}
		\label{eq::thrm2}
		W_i=\big(\exp([g_2,\cdot\,])\exp([g_3,\cdot\,])\cdots\big)w_i,
	\end{equation}
	where the generators $g_2,g_3,$ etc.\ are homogeneous polynomials in the variables $w_i$ of degree $2,3,$ etc. Here, we omit the concatenation symbol $\circ$, as it is clear from the context that we are concatenating operators. Recall that a homogeneous polynomial $p_m(\bm{w})$ of degree $m$ satisfies the following condition:
	\begin{equation}
	\label{eq::homPoly}
		p_m(\lambda\bm{w})=\lambda^m\,p_m(\bm{w}),
	\end{equation} 
	for every $\lambda\in\mathbb{R}$. Note that the symplectic map described in Eq.~\eqref{eq::anSym} maps the origin of phase space into itself, i.e., $\bm{W}(\bm{0})=\bm{0}$. The core information to be taken from these two theorems is that any analytic symplectic map of the form \eqref{eq::anSym} can be represented as an infinite product of Lie transformations with homogeneous polynomials as generators and that the truncated product is still a symplectic map, since it is a product of symplectic maps \cite{DragtFinn}.
	
	We conclude this section with some results which are necessary for the derivations in the sections to follow. The first is the Baker-Campbell-Hausdorff (BCH) formula \cite{DragtFinn}. Given two functions $g$ and $f$, the function $k$ satisfying
	\begin{subequations}
		\label{eq::BCH}
		\begin{equation}
			\exp([k,\cdot\,])=\exp([g,\cdot\,])\exp([f,\cdot\,]),
		\end{equation} 
		is given by
		\begin{equation}
			k=g+f+[g,f]/2+([g,[g,f]]+[f,[f,g]])/12+\cdots\quad.
		\end{equation}
	\end{subequations}
	If $f$ and $g$ commute, i.e., $[f,g]=0$, then we have that
	\begin{equation}
		\label{eq::BCHcommute}
		\exp([g,\cdot\,])\exp([f,\cdot\,])=\exp([g+f,\cdot\,]).
	\end{equation}
	Note that the Lie transformation associated to a constant only generates the identity operator. Thus, the inverse of a Lie transformation follows from Eq.~\eqref{eq::BCHcommute}:
	\begin{equation}
	\label{eq::LieInverseTest}
		\exp(-[f,\cdot\,])\exp([f,\cdot\,])=\exp([f,\cdot\,])\exp(-[f,\cdot\,])=I.
	\end{equation}
	From Eq.~\eqref{eq::LieInverseTest} we deduce that
	\begin{equation}
		\label{eq::LieInverse}
		\exp([f,\cdot\,])^{-1}=\exp(-[f,\cdot\,]).
	\end{equation}
	The second Lie algebraic tool of relevance is the following result, which will facilitate considerably the treatment of products of Lie transformations. Theorem 3 of Dragt and Finn \cite{DragtFinn} states that for any two Lie operators $[f,\cdot\,]$ and $[g,\cdot\,]$ we have that
	\begin{equation}
		\begin{gathered}
			\exp([f,\cdot\,])\exp([g,\cdot\,])\exp(-[f,\cdot\,])=\exp([k,\cdot\,]),\\
			k=\exp([f,\cdot\,])g.
		\end{gathered}
		\label{eq::thrm3}
	\end{equation}

	Lastly, we make a remark about the order of approximation that we will consider in this work. All the expansions of phase space variables will be performed up to third order terms, i.e., $\vert\sigma\vert\leq3$ in Eq.~\eqref{eq::vectorNorm}. This decision imposes some restrictions on the degree of the generators $g_2,g_3,\ldots$ of the Lie transformations in Eq.~\eqref{eq::thrm2}. All stems from the observation that if we consider the space $\mathcal{P}_m$ spanned by monomials of degree $m\geq 1$, then for $f\in\mathcal{P}_m$ and $g\in\mathcal{P}_n$ we have that
	\begin{equation}
		\label{eq::orderReduction}
		[f,g]\in\mathcal{P}_{m+n-2}.
	\end{equation}
	As such, if we decide to consider polynomial expansions in phase space only up to third order, then we can restrict the degree of homogeneous polynomials in Eq.~\eqref{eq::thrm2} to four. Any Lie transformation whose associated polynomial is of degree higher than four will generate terms of order higher than three and can therefore be neglected. As another consequence of property \eqref{eq::orderReduction}, Lie transformations associated with homogeneous polynomials of second degree do not change the degree of their arguments. Therefore, when applied to phase space variables they can be represented by linear transformations. We will refer to these Lie transformations as the Gaussian part of the mapping.

	\section{Lie Algebraic Formulation of Geometrical Optics in Phase Space}\label{sec::LieForm}
	Using the results in Section \ref{sec::LieTools}, we will describe how to formulate the propagation and refraction shown in Section \ref{sec::ClassicForm} in terms of Lie transformations. All the results which are to follow are calculated up to third order terms in the phase space variable expansions for rotationally symmetric optical systems.
	
	We start by proving that the Hamiltonian defined in Eq.~\eqref{eq::ham2} is the generator of the solution of free propagation in a medium of constant refractive index. 	First, note that the Hamiltonian equations \eqref{eq::HamEq} can be represented in terms of Poisson brackets:
	\begin{equation}
		\label{eq::HamEqLieForm}
		\dot{\bm{q}}=-[H,\bm{q}],\quad\dot{\bm{p}}=-[H,\bm{p}].
	\end{equation}
	Using the notation of Eqs.~\eqref{eq::wVars}, we can use the even more compact representation:
	\begin{equation}
		\label{eq::HamEqLieForm2}
		\dot{\bm{w}}=-[H,\bm{w}].
	\end{equation}
	Consider the Taylor expansion of $\bm{w}$ at the final image screen $z=z^\mathrm{f}$:
	\begin{equation}
	\label{eq::solutionTaylor}
		\bm{w}(z^\mathrm{f})=\bm{w}(z^\mathrm{i})+\sum_{m=1}^\infty\frac{1}{m!}(z^\mathrm{f}-z^\mathrm{i})^m\left.\frac{\mathrm{d}^m\bm{w}}{\mathrm{d}z^m}\right\vert_{z=z^\mathrm{i}}.
	\end{equation}
	We know from Eq.~\eqref{eq::HamEqLieForm2} that
	\begin{equation}
	\label{eq::timePoiss}
		\frac{\mathrm{d}\bm{w}}{\mathrm{d}z}=-[H,\cdot\,]\bm{w},\quad\frac{\mathrm{d}^2\bm{w}}{\mathrm{d}z^2}=-[H,\cdot\,]\frac{\mathrm{d}\bm{w}}{\mathrm{d}z}=(-[H,\cdot\,])^2\bm{w},\quad\ldots\quad,
		\frac{\mathrm{d}^m\bm{w}}{\mathrm{d}z^m}=(-[H,\cdot\,])^m\bm{w}.
	\end{equation}
	Upon substitution of Eqs.~\eqref{eq::timePoiss} into the Taylor expansion \eqref{eq::solutionTaylor}, we get the series representation of a Lie transformation as in Eq.~\eqref{eq::LieTransformation} with the Hamiltonian $H$ as generator. This means, that the mapping $\mathcal{M}$, such that $\bm{w}(z^\mathrm{f})=\mathcal{M}\big(\bm{w}(z^\mathrm{i})\big)$ and which is a solution to Eqs.~\eqref{eq::HamEqLieForm2}, is defined as 
	\begin{equation}
		\label{eq::hamGenerator}
		\mathcal{M}=\exp(-(z^\mathrm{f}-z^\mathrm{i})[H,\cdot\,]).
	\end{equation} 
	Given the result of Eq.~\eqref{eq::hamGenerator}, we have that free propagation from a standard screen at $z=z^\mathrm{i}$ to a standard screen at $z=z^\mathrm{f}$ is described by 
	\begin{subequations}
	\label{eq::hamSols}
		\begin{align}
			\exp(-(z^\mathrm{f}-z^\mathrm{i})	[H,\cdot\,])\bm{q}&=\bm{q}+(z^\mathrm{f}-z^\mathrm{i})\frac{\bm{p}}{\sqrt{n^2-\vert\bm{p}\vert^2}},\\
			\exp(-(z^\mathrm{f}-z^\mathrm{i})	[H,\cdot\,])\bm{p}&=\bm{p}.
		\end{align}
	\end{subequations}
	The solution in Eqs.~\eqref{eq::hamSols} is in agreement with Eq.~\eqref{eq::posConservation}.	Note that the Hamiltonian can be expanded as follows:
	\begin{equation}
	\label{eq::hamExpansion}
	    \begin{aligned}
		-H=\sqrt{n^2-\vert \bm{p}\vert^2}&=n-\frac{\vert\bm{p}\vert^2}{2n}-\frac{(\vert\bm{p}\vert^2)^2}{8n^3}-\cdots\\
		&=-h_0-h_2(\bm{p})-h_4(\bm{p})-\cdots.
	\end{aligned}
	\end{equation}
	Thus, the transit through a slab of thickness $d$ made of a homogeneous medium with refractive index $n$ can alternatively be written as an infinite product of Lie transformations:
	\begin{align}
		\mathcal{M}&=\exp\left(d\left[\sqrt{n^2-\vert\bm{p}\vert^2},\cdot\,\right]\right)\nonumber\\
		&=\exp(-d/(2n)[\vert\bm{p}\vert^2,\cdot\,])\exp(-d/(8n^3)[(\vert\bm{p}\vert^2)^2,\cdot\,])\cdots\,
		\label{eq::transit}\\
		&=\exp(-d[h_2,\cdot\,])\exp(-d[h_4,\cdot\,])\cdots\,\,.\nonumber
	\end{align}
	The identity in Eq.~\eqref{eq::transit} holds because the generators in the sole variable $\vert\bm{p}\vert^2$, present in the expansion \eqref{eq::hamExpansion}, commute by the property \eqref{eq::firstLie} and therefore the BCH formula \eqref{eq::BCH} is significantly simplified, see Eq.~\eqref{eq::BCHcommute}. The term $h_0$ clearly does not generate any Lie transformation other than identity. Lastly, recall that by the argument following from Eq.~\eqref{eq::orderReduction}, the linear part of the mapping $\mathcal{M}$ is entirely generated by $h_2$. The linear, or Gaussian, part of the mapping therefore reads
	\begin{equation}
	\label{eq::gaussianPropagation}
	    \mathcal{M}_\mathrm{G}=\exp(-d[h_2,\cdot\,]).
	\end{equation}
	The map in Eq.~\eqref{eq::gaussianPropagation} can be represented by a matrix:
	\begin{equation}
	\label{eq::matrixPropagation}
	    \exp(-d[h_2,\cdot\,])\bm{w}=\begin{pmatrix}
			I_{\scriptscriptstyle 2\times2} & \dfrac{d}{n}I_{\scriptscriptstyle 2\times2}\\ O_{\scriptscriptstyle 2\times2} & I_{\scriptscriptstyle 2\times2}
		\end{pmatrix}\bm{w}.
	\end{equation}
	
	We now turn our attention to the refraction mapping described in Eqs.~\eqref{eq::primed}. The optical surface is given by Eq.~\eqref{eq::rotSurf}, where terms of order higher than four will not influence the third order aberrations. Thus we only consider
	\begin{equation}
	\label{eq::surfaceEq}
		z=\zeta(\bm{q})=\beta\vert\bm{q}\vert^2+\delta(\vert\bm{q}\vert^2)^2.
	\end{equation} 
	The root transformations $\mathcal{R}_{n;\zeta},\mathcal{R}_{n';\zeta}$ are canonical as long as the considered ray lies in the Descartes sphere $\vert\bm{p}\vert^2<n^2$ and intersects the surface \cite{SaadWolf1986,Wolf2004,DragtFoundations86}. Consequently, their inverses are canonical too. Thus, the refraction mapping $\mathcal{S}_{n,n';\zeta}$ is canonical since it is a composition of canonical operators \cite{Wolf2004,DragtFoundations86}. Furthermore, under the appropriate conditions on the surface, i.e., $\zeta(\bm{0})=0$, $\nabla\zeta(\bm{0})=\bm{0}$, the phase space origin is mapped onto itself. As such, the results of Eq.~\eqref{eq::thrm2} can be applied \cite{Dragt82}. It is now a matter of finding the appropriate homogeneous polynomials which generate the mapping. We can already make a few statements. Since in this paper we will restrict ourselves to the treatment of third order aberrations by considering only third order expansion terms, any Lie transformations generated by homogeneous polynomials of degree five or higher can be neglected. Additionally, we will consider rotationally symmetric systems. This implies that the polynomials themselves can only depend on the variables $\vert\bm{p}\vert^2$, $\vert\bm{q}\vert^2$, $\bm{p}\cdot\bm{q}$ and $\bm{p}\times\bm{q}$ \cite{Dragt82,Wolf2004,Wolf88}. Since $\bm{p}\times\bm{q}$ is a preserved quantity in rotationally symmetric systems we will omit it. As such, refraction can only be associated with polynomials of even degree and we will now search for the two polynomials $g_2$, $g_4$ such that the product of the associated Lie transformations applied to our initial phase space variables reproduces the results in Eqs.~\eqref{eq::primed} up to third order, i.e.,
	\begin{equation}
	\label{eq::thirdOrderRefrLieMap}
		\bm{w}'=\exp([g_2,\cdot\,])\exp([g_4,\cdot\,])\bm{w}+\mathcal{O}(w^5),
	\end{equation}
	where $w^5$ is defined according to Eq.~\eqref{eq::vectorNorm} and the $\mathcal{O}$ terms are to be considered in each component of the vector $\bm{w}'$. It can be verified \cite{Dragt82,DragtFoundations86,SaadWolf1986} that Gaussian refraction is associated with 
	\begin{equation}
		\label{eq::gaussRefr}
		g_2(\bm{q},\bm{p})=\beta(n-n')\vert\bm{q}\vert^2,
	\end{equation} 
	where $2\beta$ is the curvature of the surface at the origin, see Eq.~\eqref{eq::surfaceEq}. The corresponding linear transformation modifies the initial variables according to
	 \begin{equation}
	 \label{eq::matrixRefraction}
		\exp([g_2,\cdot\,])\binom{\bm{q}}{\bm{p}}=\begin{pmatrix}
			I_{\scriptscriptstyle 2\times2} & O_{\scriptscriptstyle 2\times2} \\ 2\beta(n-n')I_{\scriptscriptstyle 2\times2} & I_{\scriptscriptstyle 2\times2}
		\end{pmatrix}\binom{\bm{q}}{\bm{p}}=\binom{\bm{q}}{\bm{p}+2\beta(n-n')\bm{q}},
	\end{equation}
	which corresponds to the Gaussian part in Eqs.~\eqref{eq::primed}.

	Next, let us consider the third order transverse ray aberrations as being associated with a fourth degree polynomial $g_4(\bm{q},\bm{p})$ of the general form 
	\begin{equation}
		\label{eq::poly4general}
		g_4(\bm{q},\bm{p})=A(\vert\bm{p}\vert^2)^2+B(\bm{p}\cdot\bm{q})\vert\bm{p}\vert^2+C(\bm{p}\cdot\bm{q})^2+D\vert\bm{p}\vert^2\vert\bm{q}\vert^2+E(\bm{p}\cdot\bm{q})\vert\bm{q}\vert^2+F(\vert\bm{q}\vert^2)^2.
	\end{equation}
	The generated transformation up to third order reads:
	\small
	\begin{equation}
	\label{eq::g4Action}
	    \begin{aligned}
		\binom{\bm{q}_{g_4}(\bm{q},\bm{p})}{\bm{p}_{g_4}(\bm{q},\bm{p})}
		=&\exp([g_4,\cdot\,])\binom{
			\bm{q}}{\bm{p}}=\binom{\bm{q}}{\bm{p}}+\binom{[g_4,\bm{q}]}{[g_4,\bm{p}]}+\mathcal{O}(w^5)\\
		=&	
		\binom{
			\bm{q}-4A\vert\bm{p}\vert^2\bm{p}-B\big(2(\bm{p}\cdot \bm{q})\bm{p}+\vert\bm{p}\vert^2\bm{q}\big)-2C(\bm{p}\cdot \bm{q})\bm{q}-2D\vert\bm{q}\vert^2\bm{p}-E\vert\bm{q}\vert^2\bm{q}}{
			\bm{p}+B\vert\bm{p}\vert^2\bm{p}+2C(\bm{p}\cdot \bm{q})\bm{p}+2D\vert\bm{p}\vert^2\bm{q}+E\big(2(\bm{p}\cdot\bm{q}) \bm{q}+\vert\bm{q}\vert^2\bm{p}\big)+4F\vert\bm{q}\vert^2\bm{q}}+\mathcal{O}(w^5),
	\end{aligned}
	\end{equation}
	\normalsize
	where we use only the first two terms of the exponential series of $g_4$ and $[g_4,\bm{q}]=-\partial g_4/\partial\bm{p},\,[g_4,\bm{p}]=\partial g_4/\partial\bm{q}$. In fact, any subsequent term is of order greater than three and can be omitted. We now concatenate the two transformations in Eq.~\eqref{eq::thirdOrderRefrLieMap} to find
	\begin{align}
		\exp([g_2,\cdot\,])\exp([g_4,\cdot\,])\binom{
			\bm{q} }{ \bm{p}}=&
		\exp([g_2,\cdot\,])\binom{
			\bm{q}_{g_4}(\bm{q},\bm{p})}{\bm{p}_{g_4}(\bm{q},\bm{p})}\nonumber\\
		=&
		\binom{
			\bm{q}_{g_4}(\exp([g_2,\cdot\,])\bm{q},\exp([g_2,\cdot\,])\bm{p})}{\bm{p}_{g_4}(\exp([g_2,\cdot\,])\bm{q},\exp([g_2,\cdot\,])\bm{p})}\label{eq::thirdOrderRefraction}\\
		=&\binom{
			\bm{q}'}{\bm{p}'}.\nonumber
	\end{align}
	Here we used the property of ``jumping into arguments'' of Lie transformations described in Eq.~\eqref{eq::jumpingIntoArgs}. 
	If we perform the operations in Eq.~\eqref{eq::thirdOrderRefraction} and compare with the expressions in Eqs.~\eqref{eq::primed}, then we get the following results for the coefficients $A,\ldots,F$ in Eq.~\eqref{eq::poly4general}, which are consistent with \cite{SaadWolf1986,DragtFoundations86}:
	\begin{align}
		A&=0, \quad B=0, \quad C=0,\nonumber\\
		D&=\frac{\beta}{2}\left(\frac{1}{n'}-\frac{1}{n}\right),\quad E=2\beta^2\left(1-\frac{n'}{n}\right),\label{eq::poly4Coeff}\\ 
		F&=\delta(n-n')-2\beta^3\frac{(n-n')^2}{n}\nonumber.
	\end{align}
	We have now fully determined the generators of free propagation in a homogeneous medium, see Eqs.~\eqref{eq::hamExpansion},\eqref{eq::transit} and refraction by a rotationally symmetric surface, see Eqs.~\eqref{eq::thirdOrderRefrLieMap},~\eqref{eq::gaussRefr},~\eqref{eq::poly4general} and ~\eqref{eq::poly4Coeff}. The coefficients of the generating polynomials are completely defined by the geometric parameters of the underlying system. Before concluding this section, note that in Eq.~\eqref{eq::g4Action} the coefficient $F$ of $g_4$ is the only one that does not influence the final position coordinates and $A$ does not influence the final direction coordinates.

	\section{Fundamental Refractor Element}\label{sec::refrElandSys}
	With the description of free propagation and refraction up to third order, we now aim to describe what may be the most fundamental refractive optical element: a single refractive surface. The resulting map $\mathcal{M}$ will be split into a Gaussian and a higher order part. Once we have developed the treatment for one surface, the extension to rotationally symmetric refractive optical systems will be straightforward. 
	
	We split refraction through an optical interface into three steps: propagation from the object to the interface, refraction at the interface and propagation of the refracted ray to the image plane, see Figure \ref{fig::rootTransf}. The object and image distances will be denoted by $s_\mathrm{o}$ and $s_\mathrm{i}$, respectively, and they satisfy the Gaussian imaging condition \cite{Hecht}
	\begin{equation}
	\label{eq::gaussianImaging}
		\frac{n}{s_\mathrm{o}}+\frac{n'}{s_\mathrm{i}}=\frac{n'-n}{R},
	\end{equation}
	where $R=(2\beta)^{-1}$ is the radius of curvature of the interface. By combining the results in Eqs.~\eqref{eq::hamExpansion},\eqref{eq::transit},\eqref{eq::gaussRefr},\eqref{eq::poly4Coeff}, the symplectic mapping $\mathcal{M}$ with generators up to fourth degree for a single refracting optical element reads
	\small
	\begin{multline}
		\mathcal{M}=\underbrace{\exp\left(-\frac{s_\mathrm{o}}{2n}[\vert\bm{p}\vert^2,\cdot\,]\right)\exp\left(-\frac{s_\mathrm{o}}{8n^3}[(\vert\bm{p}\vert^2)^2,\cdot\,]\right)}_{\text{propagation from object plane}}\underbrace{\exp(\beta(n-n')[\vert\bm{q}\vert^2,\cdot\,])\exp([g_4,\cdot\,])}_{\text{refraction}}\\
		\underbrace{\exp\left(-\frac{s_\mathrm{i}}{2n'}[\vert\bm{p}\vert^2,\cdot\,]\right)\exp\left(-\frac{s_\mathrm{i}}{8n'^3}[(\vert\bm{p}\vert^2)^2,\cdot\,]\right)}_{\text{propagation to image plane}},
		\label{eq::lensRefraction}
	\end{multline}
	\normalsize
	where $g_4$ is completely defined by the coefficients in Eq.~\eqref{eq::poly4Coeff}. Using Lie algebra techniques described in Section \ref{sec::LieTools}, it is possible to express Eq.~\eqref{eq::lensRefraction} in the product form
	\begin{equation}
		\mathcal{M}\approx\mathcal{M}_\mathrm{G}\exp([t_4,\cdot\,]),
		\label{eq::orderedMap}
	\end{equation}
	where $\mathcal{M}_\mathrm{G}$ contains all information regarding the Gaussian approximation and $t_4$ contains all the higher order aberration information of the complete fundamental element (up to third order transverse ray aberrations). Generators of degree five or higher are omitted, since their action on the arguments generates terms of order four or higher.  
	
	By considering only the contributions of the Gaussian part $\mathcal{M}_\mathrm{G}$ in Eq.~\eqref{eq::orderedMap}, we recover the traditional matrix formulation for refraction of rays at an interface \cite{Hecht}, where
	\begin{equation}
		\mathcal{M}_\mathrm{G}=\exp\left(-\frac{s_\mathrm{o}}{2n}[\vert\bm{p}\vert^2,\cdot\,]\right)\exp(\beta(n-n')[\vert\bm{q}\vert^2,\cdot\,])\exp\left(-\frac{s_\mathrm{i}}{2n'}[\vert\bm{p}\vert^2,\cdot\,]\right).\label{eq::GaussianInterfaceOperators}
	\end{equation}
	Since matrices do not commute and do not ``jump into arguments'' like Lie transformations, their order needs to be inverted compared to the associated Lie transformations product. Using the matrix representations given in Eq.~\eqref{eq::matrixPropagation} and Eq.~\eqref{eq::matrixRefraction}, we rewrite the operator in Eq.~\eqref{eq::GaussianInterfaceOperators} as 
	\begin{equation}
		\mathcal{M}_\mathrm{G}=\begin{pmatrix}
			I_{\scriptscriptstyle 2\times2} & \dfrac{s_\mathrm{i}}{n'}I_{\scriptscriptstyle 2\times2}\\ O_{\scriptscriptstyle 2\times2} & I_{\scriptscriptstyle 2\times2}
		\end{pmatrix}
		\begin{pmatrix}
			I_{\scriptscriptstyle 2\times2} & O_{\scriptscriptstyle 2\times2} \\ 2\beta(n-n')I_{\scriptscriptstyle 2\times2} & I_{\scriptscriptstyle 2\times2}
		\end{pmatrix}
		\begin{pmatrix}
			I_{\scriptscriptstyle 2\times2} & \dfrac{s_\mathrm{o}}{n}I_{\scriptscriptstyle 2\times2}\\ O_{\scriptscriptstyle 2\times2} &I_{\scriptscriptstyle 2\times2} 
		\end{pmatrix}.
		\label{eq::gaussianInterfaceMatrix3}
	\end{equation}
	Let us introduce the dioptric power of a surface $\mathcal{D}=(n'-n)/R$, the magnification $m=-ns_\mathrm{i}/(n's_\mathrm{o})$ and recall that $2\beta=1/R$, where $R$ is the radius of curvature of the surface \cite{Hecht}. Then the Gaussian imaging condition \eqref{eq::gaussianImaging} turns Eq.~\eqref{eq::gaussianInterfaceMatrix3} into
	\begin{equation}
		\mathcal{M}_\mathrm{G}=\begin{pmatrix}
			-\dfrac{ns_\mathrm{i}}{n's_\mathrm{o}}I_{\scriptscriptstyle 2\times2} & O_{\scriptscriptstyle 2\times2}\\
			-\mathcal{D}\,I_{\scriptscriptstyle 2\times2} & -\dfrac{n's_\mathrm{o}}{ns_\mathrm{i}}I_{\scriptscriptstyle 2\times2}
		\end{pmatrix}=\begin{pmatrix}
			mI_{\scriptscriptstyle 2\times2} & O_{\scriptscriptstyle 2\times2}\\
			-\mathcal{D}\,I_{\scriptscriptstyle 2\times2} & \dfrac{1}{m}I_{\scriptscriptstyle 2\times2}
		\end{pmatrix}.
		\label{eq::GaussianInterface}
	\end{equation}
	The matrix in Eq.~\eqref{eq::GaussianInterface} represents the traditional matrix formulation of Gaussian optics \cite{Hecht}.
	
	To group the Gaussian parts together, we have to apply some manipulations to the original mapping $\mathcal{M}$. We use the result of Eq.~\eqref{eq::LieInverse} to insert identity operators. Thus, we ``create'' triplets which enable the use of Eq.~\eqref{eq::thrm3}. We simplify the notation by substituting the generators of propagation to the object plane  $-s_\mathrm{o}[h_2,\cdot\,]$, $-s_\mathrm{o}[h_4,\cdot\,]$, the generators of refraction  $[g_2,\cdot\,]$, $[g_4,\cdot\,]$ and the generators of propagation to the image plane $-s_\mathrm{i}[h'_2,\cdot\,],-s_\mathrm{i}[h'_4,\cdot\,]$. Here, the subscript represents the degree of the respective homogeneous polynomial and the $h'$ polynomials take into consideration the image space refractive index $n'$, cf. Eq.~\eqref{eq::hamExpansion}. We rewrite Eq.~\eqref{eq::lensRefraction} as 
	\begin{align*}
		\mathcal{M}=\,&\exp\left(-s_\mathrm{o}[h_2,\cdot\,]\right)\exp\left(-s_\mathrm{o}[h_4,\cdot\,]\right)\\
		&\exp([g_2,\cdot\,])\exp([g_4,\cdot\,])\\	&\exp\left(-s_\mathrm{i}[h'_2,\cdot\,]\right)\exp\left(-s_\mathrm{i}[h'_4,\cdot\,]\right)\\
		=\,&\exp\left(-s_\mathrm{o}[h_2,\cdot\,]\right)\exp([g_2,\cdot\,])\exp\left(-s_\mathrm{i}[h'_2,\cdot\,]\right)\\
		&\exp\left(s_\mathrm{i}[h'_2,\cdot\,]\right)\exp(-[g_2,\cdot\,])\exp(-s_\mathrm{o}[h_4,\cdot\,])\\
		&\exp([g_2,\cdot\,])\exp([g_4,\cdot\,])\\	&\exp\left(-s_\mathrm{i}[h'_2,\cdot\,]\right)\exp\left(-s_\mathrm{i}[h'_4,\cdot\,]\right)\\
		=\,&\mathcal{M}_\mathrm{G}\\
		&\exp\left(s_\mathrm{i}[h'_2,\cdot\,]\right)\exp(-[g_2,\cdot\,])\exp(-s_\mathrm{o}[h_4,\cdot\,])\\
		&\exp([g_2,\cdot\,])\exp([g_4,\cdot\,])\\
		&	\exp\left(-s_\mathrm{i}[h'_2,\cdot\,]\right)\exp\left(-s_\mathrm{i}[h'_4,\cdot\,]\right).\numberthis\label{eq::modMapping1}
	\end{align*}
	We rewrite the mapping in Eq.~\eqref{eq::modMapping1} and insert one more time an identity operator of the form $I=\exp\left(-s_\mathrm{i}[h'_2,\cdot\,]\right)\exp\left(s_\mathrm{i}[h'_2,\cdot\,]\right)$:
	\begin{align*}
		\mathcal{M}=\,&\mathcal{M}_\mathrm{G}\\
		&\exp\left(s_\mathrm{i}[h'_2,\cdot\,]\right)\exp(-[g_2,\cdot\,])\exp(-s_\mathrm{o}[h_4,\cdot\,])
		\exp([g_2,\cdot\,])\exp\left(-s_\mathrm{i}[h'_2,\cdot\,]\right)\\
		&\exp\left(s_\mathrm{i}[h'_2,\cdot\,]\right)\exp([g_4,\cdot\,])	\exp\left(-s_\mathrm{i}[h'_2,\cdot\,]\right)\\
		&\exp\left(-s_\mathrm{i}[h'_4,\cdot\,]\right).\numberthis\label{eq::modMapping2}
	\end{align*}
	In Eq.~\eqref{eq::modMapping2} there are multiple instances in which we can apply the result given in Eq.~\eqref{eq::thrm3}: twice in the second line and once in the third one. This leads to
	\begin{equation}
		\mathcal{M}=\,\mathcal{M}_\mathrm{G}\exp(-s_\mathrm{o}[\widetilde{h}_4,\cdot\,])\exp([\widetilde{g}_4,\cdot\,])\exp(-s_\mathrm{i}[h'_4,\cdot\,])\label{eq::modMapping3},
	\end{equation}
	where $\widetilde{h}_4$ and $\widetilde{g}_4$ are given by
	\begin{equation}
		\label{eq::transformedPoly}
		\widetilde{h}_4=-s_\mathrm{o}\exp\left(s_\mathrm{i}[h'_2,\cdot\,]\right)\exp(-[g_2,\cdot\,])h_4,\quad \widetilde{g}_4=\exp(s_\mathrm{i}[h'_2,\cdot\,])g_4.
	\end{equation}
 	Recall that Lie transformations associated with homogeneous polynomials of second degree do not change the degree of their arguments and hence $\widetilde{h}_4,\,\widetilde{g}_4$ are of degree 4. The last necessary step, in order to derive the desired expression Eq.~\eqref{eq::orderedMap}, is to apply the BCH formula \eqref{eq::BCH} to combine the three fourth degree Lie transformations into one. This step is straightforward, since Poisson brackets between fourth degree generators are generating homogeneous polynomials of degree six or higher and can therefore be neglected, see the discussion at the end of Section \ref{sec::LieTools}. It follows from Eq.~\eqref{eq::modMapping3}, that we can write $t_4$ in Eq.~\eqref{eq::orderedMap} as
	\begin{equation}
	\label{eq::t4}
		t_4=-s_\mathrm{o}\widetilde{h}_4+\widetilde{g}_4-s_\mathrm{i}h'_4.
	\end{equation} 
	After explicitly performing the computations given in Eqs.~\eqref{eq::transformedPoly} and Eq.~\eqref{eq::t4}, we denote with $A^\ast,\ldots,F^\ast$ the coefficients of $t_4$, analogously to the representation of $g_4$ in Eq.~\eqref{eq::poly4general} and find 
	\begin{equation}
		\begin{cases}
			A^\ast=-\dfrac{s_\mathrm{o}}{8n^3}m^4+\dfrac{s_\mathrm{i}^2}{n'^2}D-\dfrac{s_\mathrm{i}^3}{n'^3}E+\dfrac{s_\mathrm{i}^4}{n'^4}F-\dfrac{s_\mathrm{i}}{8n'^3},\\
			B^\ast=-\dfrac{s_\mathrm{o}\mathcal{D}}{2n^3}m^3-2\dfrac{s_\mathrm{i}}{n'}D+3\dfrac{s_\mathrm{i}^2}{n'^2}E-4\dfrac{s_\mathrm{i}^3}{n'^3}F,\\
			C^\ast=-\dfrac{s_\mathrm{o}\mathcal{D}^2}{2n^3}m^2-2\dfrac{s_\mathrm{i}}{n'}E+4\dfrac{s_\mathrm{i}^2}{n'^2}F,\\
			D^\ast=-\dfrac{s_\mathrm{o}\mathcal{D}^2}{4n^3}m^2+D-\dfrac{s_\mathrm{i}}{n'}E+2\dfrac{s_\mathrm{i}^2}{n'^2}F,\\
			E^\ast=-\dfrac{s_\mathrm{o}\mathcal{D}^3}{2n^3}m+E-4\dfrac{s_\mathrm{i}}{n'}F,\\
			F^\ast=-\dfrac{s_\mathrm{o}\mathcal{D}^4}{8n^3}+F,
		\end{cases}
		\label{eq::InterfaceAberrPolyCoeffGENERAL}
	\end{equation}
	where $D,E,F$ are defined in Eqs.~\eqref{eq::poly4Coeff}. It is now possible to represent our fundamental element in the form Eq.~\eqref{eq::orderedMap} with the Gaussian part described by Eq.~\eqref{eq::GaussianInterfaceOperators}, or equivalently Eq.~\eqref{eq::GaussianInterface}, and the fourth degree polynomial defined by its coefficients in Eqs.~\eqref{eq::InterfaceAberrPolyCoeffGENERAL}. 
	
	We conclude this section with a brief comment about higher order aberrations. The methods described in Section \ref{sec::LieForm} and Section \ref{sec::refrElandSys} allow for further investigation of the aberrations of optical interfaces. The derivation of generating polynomials of order six and higher, which correspond to transverse ray aberrations of order five and higher, follows the same steps as described so far. It needs to be noted that the composition of operators via the BCH formula becomes increasingly more complicated as lower order aberrations compose into higher order contributions in addition to the intrinsic aberrations of the surface. For instance, composing the generators of third order aberrations will lead to additional contributions to the generators of fifth order aberrations. The necessary machinery needed for these computations is however already contained in the mathematical framework of the Lie algebraic method.
	
	\section{Optical Systems}\label{sec::OptSys}
	With the results of Section \ref{sec::refrElandSys}, it is possible to concatenate multiple fundamental elements into a complete optical system. The necessary tools have already been presented and used in the previous sections.
	
	A rotationally symmetric refractive optical system is simply a composition of fundamental elements as described in Section \ref{sec::refrElandSys}. We therefore consider an optical system with $k$ interfaces. Each interface can be described up to fourth order terms by its coefficients $\beta_j,\delta_j,j=1,\dots,k$ according to
	\begin{equation}
	\label{eq::surfaceEqSystem}
		z=\zeta_j(\bm{q})=\beta_j\vert\bm{q}\vert^2+\delta_j(\vert\bm{q}\vert^2)^2,\quad j=1,\dots,k.
	\end{equation}
	With the surface equation \eqref{eq::surfaceEqSystem} at the vertex of each surface, we can determine the paraxial quantities. At each interface $j$ the Gaussian imaging conditions are satisfied
	\begin{equation}
	\label{eq::gaussImagingCond}
		\frac{n_j}{s_{\mathrm{o},j}}+\frac{n_{j+1}}{s_{\mathrm{i},j}}=\frac{n_{j+1}-n_j}{R_j}.
	\end{equation}
	In Eq.~\eqref{eq::gaussImagingCond} it holds that $2\beta_j=R_j^{-1}$. The additional quantities which are necessary for the calculations are:
	\begin{equation}
		\label{eq::gaussQuant}
		m_j=-\frac{n_js_{\mathrm{i},j}}{n_{j+1}s_{\mathrm{o},j}},\quad
		s_{\mathrm{o},j+1}=d_j-s_{\mathrm{i},j},\quad
		\mathcal{D}_j=\frac{n_{j+1}-n_j}{R_j},
	\end{equation}
	where $m_j$ is the magnification of the optical surface $j$, $\mathcal{D}_j$ its dioptric power, $c_j=1/R_j$ is its curvature and $d_j$ is the distance between interface $j$ and interface $j+1$ along the optical axis. As previously shown, we can derive a symplectic mapping which describes ray propagation up to operators generated by fourth degree polynomials. We divide the contribution of each interface $j$ into a second degree, $\mathcal{M}_{\mathrm{G}_j}$, and fourth degree, $\exp([\sigma_j,\cdot\,])$, contribution. Here, $\sigma_j$ is a fourth degree polynomial whose coefficients are completely defined by the geometry of the interface $j$ and the system according to Eqs.~\eqref{eq::InterfaceAberrPolyCoeffGENERAL}. Let us denote with $\mathcal{M}$ the symplectic map describing ray propagation through our optical system up to fourth order, then we have 
	\begin{equation}
		\mathcal{M}=\mathcal{M}_{\mathrm{G}_1}\exp([\sigma_1,\cdot\,])\mathcal{M}_{\mathrm{G}_2}\exp([\sigma_2,\cdot\,])\cdots\mathcal{M}_{\mathrm{G}_{k-1}}\exp([\sigma_{k-1},\cdot\,])\mathcal{M}_{\mathrm{G}_k}\exp([\sigma_k,\cdot\,]).\label{eq::completeSystem}
	\end{equation}
	Similarly to previous discussions, we want all second degree components, which describe the paraxial regime, to be grouped together. We introduce the notation
	\begin{equation}
	\label{eq::gaussianPartNotation}
		\mathcal{M}_{\mathrm{G}_{1\to j}}=\mathcal{M}_{\mathrm{G}_1}\cdots\mathcal{M}_{\mathrm{G}_j},\quad j=2,\dots,k.
	\end{equation}
	The operator describing the Gaussian propagation of the complete system will be denoted by $\mathcal{M}_{\mathrm{G}_{1\to k}}$. Using the notation introduced in Eq.~\eqref{eq::gaussianPartNotation}, the following holds true:
	\begin{multline}
		\mathcal{M}=\mathcal{M}_{\mathrm{G}_{1\to k}}(\mathcal{M}_{\mathrm{G}_{2\to k}})^{-1}\exp([\sigma_1,\cdot\,])\mathcal{M}_{\mathrm{G}_{2\to k}}\\(\mathcal{M}_{\mathrm{G}_{3\to k}})^{-1}\exp([\sigma_2,\cdot\,])\mathcal{M}_{\mathrm{G}_{3\to k}}(\mathcal{M}_{\mathrm{G}_{4\to k}})^{-1}\exp([\sigma_3,\cdot\,])\cdots
		\mathcal{M}_{\mathrm{G}_k}\exp([\sigma_k,\cdot\,]).\label{eq::completeSystemMod1}
	\end{multline}
	The mappings in Eq.~\eqref{eq::completeSystem} and Eq.~\eqref{eq::completeSystemMod1} are equivalent, but in Eq.~\eqref{eq::completeSystemMod1} it is possible to apply multiple times the identity given in Eq.~\eqref{eq::thrm3}. We define $\sigma_j^\ast$ such that
	\begin{equation}
		\exp([\sigma_j^\ast,\cdot\,])=(\mathcal{M}_{\mathrm{G}_{j+1\to k}})^{-1}\exp([\sigma_j,\cdot\,])\mathcal{M}_{\mathrm{G}_{j+1\to k}}\quad j=1,\ldots,k-1.\label{eq::sigmaStar}
	\end{equation}
	The mapping $\mathcal{M}$ can therefore be rewritten as
	\begin{equation}
		\mathcal{M}=\mathcal{M}_{\mathrm{G}_{1\to k}}\exp([\sigma_1^\ast,\cdot\,])\exp([\sigma_2^\ast,\cdot\,])\cdots\exp([\sigma_{k-1}^\ast,\cdot\,])\exp([\sigma_k,\cdot\,]).
		\label{eq::completeSystemMod2}
	\end{equation}
	A computationally convenient procedure to compute the $\sigma^\ast_j$ is to consider that $\mathcal{M}_{\mathrm{G}_{j+1\to k}}$ can be treated both as a Lie transformation and as a matrix --- beware of the inversion of concatenation order when composing the matrices instead of the Lie transformations. Applying the result in Eq.~\eqref{eq::thrm3} to the definition in Eq.~\eqref{eq::sigmaStar} gives
	\begin{equation}
		\label{eq::sigmaStarMod1}
		\sigma_j^\ast=(\mathcal{M}_{\mathrm{G}_{j+1\to k}})^{-1}\sigma_j.
	\end{equation}
	The function $\sigma_j$ is a fourth degree homogeneous polynomial in $\bm{q}$ and $\bm{p}$, i.e., $\sigma_j(\bm{q},\bm{p})$. Using the property of ``jumping into arguments'' in Eq.~\eqref{eq::jumpingIntoArgs}, we rewrite Eq.~\eqref{eq::sigmaStarMod1} as
	\begin{equation}
	\label{eq::sigmaStarCalc}
		\sigma_j^\ast(\bm{q},\bm{p})=\sigma_j((\mathcal{M}_{\mathrm{G}_{j+1\to k}})^{-1}\bm{q},(\mathcal{M}_{\mathrm{G}_{j+1\to k}})^{-1}\bm{p}).
	\end{equation} 
	The action of $(\mathcal{M}_{\mathrm{G}_{j+1\to k}})^{-1}$ on the phase space variables in Eq.~\eqref{eq::sigmaStarCalc} can be described via the associated matrix and finding the coefficients of $\sigma_j^\ast(\bm{q},\bm{p})$ reduces to expanding and grouping terms in the RHS of Eq.~\eqref{eq::sigmaStarCalc}. All the $\sigma^\ast_j$ and $\sigma_k$ are homogeneous polynomials of fourth degree since the Gaussian parts of the mapping do not change the degree of their arguments. We claim that these fourth degree polynomials contain the necessary information regarding the Seidel sums, as will be proven in the coming section, and therefore the expressions in Eq.~\eqref{eq::sigmaStar} and Eq.~\eqref{eq::completeSystemMod2} tell us that the aberrations of each fundamental element are propagated through the system via the Gaussian mappings of the subsequent elements.

	We conclude this section by stating that the symplectic map $\mathcal{M}$ given in Eq.~\eqref{eq::completeSystemMod2} can be approximated up to fourth degree generators by 
	\begin{equation}
		\mathcal{M}\approx\mathcal{M}_{\mathrm{G}_{1\to k}}\exp([\tau,\cdot\,]).\label{eq::completeSystemFinal}
	\end{equation}
	With the BCH formula \eqref{eq::BCH} and the fact that higher degree polynomials can be neglected we get
	\begin{equation}
		\tau=\sum_{j=1}^{k-1}\sigma_j^\ast+\sigma_k.\label{eq::tau}
	\end{equation}

	A physical interpretation to why the form in Eq.~\eqref{eq::completeSystemFinal} might be desirable is that the Gaussian part $\mathcal{M}_{\mathrm{G}_{1\to k}}$ acts first on both the phase space coordinates and subsequently also on the Lie transformations by the ``jumping into arguments'' property. At this point we are left with a mapping defined on the Gaussian image space variables, which applies aberrations to these coordinates. To summarize, we first propagate through the complete system to the Gaussian image space via $\mathcal{M}_{\mathrm{G}_{1\to k}}$ and then aberrate. Since aberrations are traditionally given with respect to Gaussian coordinates \cite{Welford1986,Braat2019}, the ordering in Eq.~\eqref{eq::completeSystemFinal} comes natural.
	
	\section{Lie Transformations and Seidel Sums}\label{sec::LieToSeid}
	The performance of an optical system is often investigated by computing its Seidel coefficients \cite{Braat2019,Welford1986}. These coefficients completely describe the third order transverse ray aberrations of monochromatic light through a rotationally symmetric refractive system \cite{Braat2019,Welford1986}. In this section, we will show the relation between the coefficients of the fourth degree polynomial $\tau$ derived in Eq.~\eqref{eq::tau} and the Seidel aberration coefficients. Furthermore, it will be shown how the property of summing Seidel contributions over all interfaces is a direct consequence of the described mathematical framework.
	
	We start by recalling, that the wavefront aberration expansion is traditionally given in exit pupil and Gaussian image position coordinates \cite{Welford1986,Braat2019}. Transverse ray aberration in the Lie approach is instead given in terms of the Gaussian phase space coordinates. We therefore need to be able to express the Gaussian momentum coordinates in terms of the exit pupil coordinates $\bm{q}'$ and the Gaussian image position coordinates $\bm{q}_\mathrm{G}$. Up to first order, we can express the Gaussian momentum coordinate $\bm{p}_\mathrm{G}$ as follows:
	\begin{equation}
		\bm{p}_\mathrm{G}=\frac{n}{R_\mathrm{S}}(\bm{q}_\mathrm{G}-\bm{q}').
		\label{eq::momentumToPupil}
	\end{equation}
	Here, $n$ and $R_\mathrm{S}$ refer to the refractive index and exit pupil distance with respect to the last image plane, respectively. This means that $n=n_{k+1}$. Since we are interested in the third order transverse ray aberrations, we want to look at the difference between the Gaussian phase space variables at the image plane (denoted by the subscript G) and the third order phase space variables at the image plane. The full third order aberration expansion reads
	
	\begin{align}
		\Delta\bm{q}^{(3)}=&\exp([\tau(\bm{q}_\mathrm{G},\bm{p}_\mathrm{G})])\bm{q}_\mathrm{G}-\bm{q}_\mathrm{G}+\mathcal{O}(w^5)\nonumber\\
		=&\,[\tau,\bm{q}_\mathrm{G}]+\mathcal{O}(w^5)\nonumber\\
		=&-4A_{\tau}\vert\bm{p}_\mathrm{G}\vert^2\bm{p}_\mathrm{G}-B_{\tau}\big(2(\bm{p}_\mathrm{G}\cdot\bm{q}_\mathrm{G})\bm{p}_\mathrm{G}+\vert\bm{p}_\mathrm{G}\vert^2\bm{q}_\mathrm{G}\big)\label{eq::thirdOrderSeidel}\\
		&-2C_{\tau}(\bm{p}_\mathrm{G}\cdot\bm{q}_\mathrm{G})\bm{q}_\mathrm{G}-2D_{\tau}\vert\bm{q}_\mathrm{G}\vert^2\bm{p}_\mathrm{G}-E_{\tau}\vert\bm{q}_\mathrm{G}\vert^2\bm{q}_\mathrm{G}+\mathcal{O}(w^5)\nonumber,
	\end{align}
	where $A_{\tau},\ldots,E_{\tau}$ are the coefficients of $\tau$ as defined in Eq.~\eqref{eq::tau}. As in Eq.~\eqref{eq::g4Action}, we notice that $F_\tau$ does not influence the position coordinates.
	Substituting the identity \eqref{eq::momentumToPupil} into Eq.~\eqref{eq::thirdOrderSeidel} will yield an expression dependent on exit pupil and Gaussian image coordinates which is derived from the Lie approach:
	\small
	\begin{equation}
		\begin{aligned}
			\Delta\bm{q}^{(3)}=\,&\left(4\frac{n^3}{R_\mathrm{S}^3}A_{\tau}\right)&\vert\bm{q}'\vert^2\bm{q}'+\\
			&\left(-8\frac{n^3}{R_\mathrm{S}^3}A_{\tau}-2\frac{n^2}{R_\mathrm{S}^2}B_{\tau}\right)&(\bm{q}'\cdot\bm{q}_\mathrm{G})\bm{q}'+\\
			&\left(4\frac{n^3}{R_\mathrm{S}^3}A_{\tau}+2\frac{n^2}{R_\mathrm{S}^2}B_{\tau}+2\frac{n}{R_\mathrm{S}}D_{\tau}\right)&\vert\bm{q}_\mathrm{G}\vert^2\bm{q}'+\\
			&\left(-4\frac{n^3}{R_\mathrm{S}^3}A_{\tau}-\frac{n^2}{R_\mathrm{S}^2}B_{\tau}\right)&\vert\bm{q}'\vert^2\bm{q}_\mathrm{G}+\\
			&\left(8\frac{n^3}{R_\mathrm{S}^3}A_{\tau}+4\frac{n^2}{R_\mathrm{S}^2}B_{\tau}+2\frac{n}{R_\mathrm{S}}C_{\tau}\right)&(\bm{q}'\cdot\bm{q}_\mathrm{G})\bm{q}_\mathrm{G}+\\
			&\left(-4\frac{n^3}{R_\mathrm{S}^3}A_{\tau}-3\frac{n^2}{R_\mathrm{S}^2}B_{\tau}-2\frac{n}{R_\mathrm{S}}C_{\tau}-2\frac{n}{R_\mathrm{S}}D_{\tau}-E_{\tau}\right)&\vert\bm{q}_\mathrm{G}\vert^2\bm{q}_\mathrm{G}.
		\end{aligned}
		\label{eq::lieTransverseAberr}
	\end{equation}
	\normalsize
	The traditional fourth-order representation of the wavefront aberration with Seidel sums $S_\text{I},\ldots,S_\text{V}$ reads \cite{Welford1986,Braat2019}
	\begin{multline}
		W(\hat{\rho},\varphi;\hat{\eta})=\,\frac{1}{8}S_\text{I}\hat{\rho}^4+\frac{1}{2}S_{\text{II}}\hat{\eta}\hat{\rho}^3\cos\varphi+\frac{1}{2}S_{\text{III}}\hat{\eta}^2\hat{\rho}^2\cos^2\varphi\\
		+\frac{1}{4}(S_{\text{III}}+S_{\text{IV}})\hat{\eta}^2\hat{\rho}^2+\frac{1}{2}S_\text{V}\hat{\eta}^3\hat{\rho}\cos\varphi,
		\label{eq::generalSeidelAberr}
	\end{multline} 
	where $\hat{\rho}=(x^2+y^2)/\rho_{\text{max}}$ are the normalized pupil coordinates, $\hat{\eta}=\eta/\eta_{\text{max}}$ is the normalized image height and $\hat{\rho}\cos\varphi=y/\rho_{\text{max}}$. Here $\eta$ is the image height and $\eta_{\text{max}}$ is the maximum image height. Lastly, $\rho_{\text{max}}$ is the exit pupil radius. It is possible to relate wavefront aberrations to transverse ray aberrations \cite{Welford1986,Braat2019} through
	\begin{equation}
	    \label{eq::transverseAberr}
		\Delta\bm{q}=-\frac{R_\mathrm{S}}{n}\nabla_{(x,y)}W,
	\end{equation}
	where $\nabla_{(x,y)}$ represents the gradient in the $(x,y)$ coordinates and $R_\mathrm{S}$ is the radius of the reference sphere, or equivalently the exit pupil distance with respect to the last image plane. If we substitute Eq.~\eqref{eq::generalSeidelAberr} in Eq.~\eqref{eq::transverseAberr}, we have up to third order
	\begin{multline}
		\Delta\bm{q}^{(3)}=-\frac{R_\mathrm{S}}{n}\left[\frac{1}{2}\frac{S_\text{I}}{\rho^4_{\text{max}}}(x^2+y^2)\binom{
			x }{ y}+
		\frac{1}{2}\frac{S_\text{II}}{\rho_{\text{max}}^3\eta_{\text{max}}}\left((x^2+y^2)\binom{
			0 }{ \eta}+2y\eta\binom{
			x }{ y}\right)+\right.\\
		\left.\frac{S_\text{III}}{\rho_{\text{max}}^2\eta_{\text{max}}^2}y\eta\binom{0}{\eta}+
		\frac{1}{2}\frac{S_\text{III}+S_\text{IV}}{\rho_{\text{max}}^2\eta_{\text{max}}^2}\eta^2\binom{	x }{ y}+
		\frac{1}{2}\frac{S_\text{V}}{\rho_{\text{max}}\eta_{\text{max}}^3}\eta^2\binom{	0 }{ \eta}\right].
		\label{eq::seidelTransTraditional}
	\end{multline}
	Our variables in the Lie approach are $\bm{q}'=(x\,\, y)^T$ and $\bm{q}_\mathrm{G}=(0\,\,\eta)^T$. We therefore rewrite Eq.~\eqref{eq::seidelTransTraditional} as
	\begin{multline}
		\Delta\bm{q}^{(3)}=-\frac{R_\mathrm{S}}{n}\left[\frac{1}{2}\frac{S_\text{I}}{\rho^4_{\text{max}}}\vert\bm{q}'\vert^2\bm{q}'+
		\frac{1}{2}\frac{S_\text{II}}{\rho_{\text{max}}^3\eta_{\text{max}}}\left(\vert\bm{q}'\vert^2\bm{q}_\mathrm{G}+2(\bm{q}'\cdot\bm{q}_\mathrm{G})\bm{q}'\right)+\right.\\
		\left.\frac{S_\text{III}}{\rho_{\text{max}}^2\eta_{\text{max}}^2}(\bm{q}'\cdot\bm{q}_\mathrm{G})\bm{q}_\mathrm{G}+
		\frac{1}{2}\frac{S_\text{III}+S_\text{IV}}{\rho_{\text{max}}^2\eta_{\text{max}}^2}\vert\bm{q}_\mathrm{G}\vert^2\bm{q}'+
		\frac{1}{2}\frac{S_\text{V}}{\rho_{\text{max}}\eta_{\text{max}}^3}\vert\bm{q}_\mathrm{G}\vert^2\bm{q}_\mathrm{G}\right].
		\label{eq::seidelTransverseAberr}
	\end{multline}
	Comparing each term in Eq.~\eqref{eq::lieTransverseAberr} and Eq.~\eqref{eq::seidelTransverseAberr} yields
	\small
	\begin{equation}
		\begin{dcases}
			-\frac{R_\mathrm{S}}{n}\frac{1}{2}\frac{S_\text{I}}{\rho^4_{\text{max}}}=4\frac{n^3}{R_\mathrm{S}^3}A_{\tau},\\
			-\frac{R_\mathrm{S}}{n}\frac{S_\text{II}}{\rho_{\text{max}}^3\eta_{\text{max}}}=-8\frac{n^3}{R_\mathrm{S}^3}A_{\tau}-2\frac{n^2}{R_\mathrm{S}^2}B_{\tau},\\
			-\frac{R_\mathrm{S}}{n}\frac{S_\text{III}}{\rho_{\text{max}}^2\eta_{\text{max}}^2}=8\frac{n^3}{R_\mathrm{S}^3}A_{\tau}+4\frac{n^2}{R_\mathrm{S}^2}B_{\tau}+2\frac{n}{R_\mathrm{S}}C_{\tau},\\
			-\frac{R_\mathrm{S}}{n}\frac{1}{2}\frac{S_\text{III}+S_\text{IV}}{\rho_{\text{max}}^2\eta_{\text{max}}^2}=4\frac{n^3}{R_\mathrm{S}^3}A_{\tau}+2\frac{n^2}{R_\mathrm{S}^2}B_{\tau}+2\frac{n}{R_\mathrm{S}}D_{\tau},\\
			-\frac{R_\mathrm{S}}{n}\frac{1}{2}\frac{S_\text{V}}{\rho_{\text{max}}\eta_{\text{max}}^3}=-4\frac{n^3}{R_\mathrm{S}^3}A_{\tau}-3\frac{n^2}{R_\mathrm{S}^2}B_{\tau}-2\frac{n}{R_\mathrm{S}}C_{\tau}-2\frac{n}{R_\mathrm{S}}D_{\tau}-E_{\tau}.
		\end{dcases}
		\label{eq::LieToSeidel}
	\end{equation}
	\normalsize
	The system \eqref{eq::LieToSeidel} gives the Seidel sums in terms of the coefficients of the generator $\tau$ in the Lie algebraic description.
	
	An advantageous property of Seidel coefficients is that they can be represented as sums of interface contributions \cite{Welford1986,Braat2019}. It is therefore possible to distinguish the effect of each surface on the Seidel coefficient of the complete system and determine which interfaces need to be modified in order to reduce the aberrations of the complete system. Modifying one surface will then leave the contribution of the others unchanged. By definition of $\tau$ in Eq.~\eqref{eq::tau}, we can write
	\begin{equation}
		A_\tau=\sum_{j=1}^{k-1}A_{\sigma_j^\ast}+A_{\sigma_k}\label{eq::tauCoeffs},
	\end{equation} 
	and analogously to Eq.~\eqref{eq::tauCoeffs} we compute the other coefficients $B_\tau,\ldots,F_\tau$ of the polynomial $\tau$. The identities \eqref{eq::LieToSeidel} give linear relations between the Seidel sums of the system and the coefficients of the polynomial $\tau$. This implies that if we use the identities \eqref{eq::LieToSeidel} with the coefficients of $\sigma_j^\ast$ or $\sigma_k$, then we are actually computing the contributions of the individual surfaces to the respective Seidel sum. The property of adding interface contributions is therefore preserved by the Lie approach purely by mathematical argumentation --- mainly the BCH formula applied in Eq.~\eqref{eq::tau} --- and without the need of physical or geometric justifications, see \cite{Welford1986, Braat2019}.
	
	To summarize, in order to derive the Seidel sums and the surface contributions of a rotationally symmetric optical system, it is necessary to first determine all Gaussian quantities \eqref{eq::gaussQuant} of the system with the help of the Gaussian imaging conditions \eqref{eq::gaussImagingCond}. It is then possible to compute the coefficients of the fourth order generators for each interface using Eqs.~\eqref{eq::InterfaceAberrPolyCoeffGENERAL}. Once the system is described by its corresponding map as in Eq.~\eqref{eq::completeSystem}, we can proceed in modifying it into Eq.~\eqref{eq::completeSystemMod2}. In this form, it is possible to extract the third order aberration information of each interface and of the complete system according to Eqs.~\eqref{eq::LieToSeidel}.
	
	\section{Examples}\label{sec::Examples}
	In this section, we compute the Seidel sums for three examples, where we applied the results given in previous sections. The geometric data of the optical systems and the respective Seidel sums have been taken from \cite{FundOptDesign}.
	
\begin{figure}[htbp!]
		\centering
		\includegraphics[keepaspectratio, width=5cm]{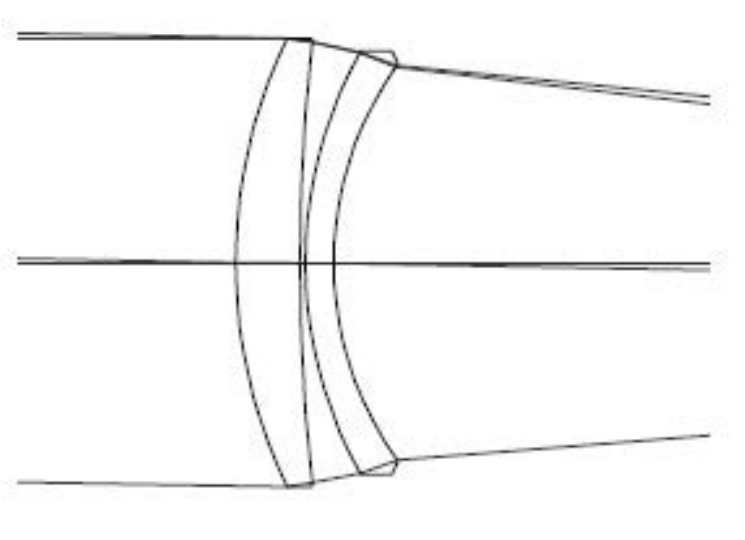}
		\caption{Gaussian doublet (Fig. 8.11, \cite{FundOptDesign}).}
		\label{fig::gaussianDoublet}
	\end{figure}

	As a first example of a fully refractive system we take a Gaussian doublet, see Figure \ref{fig::gaussianDoublet}. The necessary information is taken from \cite[p. 180]{FundOptDesign}. 
	\noindent In order to have all parameters for the expressions \eqref{eq::LieToSeidel}, we need to track the exit pupil quantities using the Gaussian part of our mapping. The Seidel sums are listed in Table \ref{tbl::gaussianDoublet}.
	\begin{table}[htbp!]
		\centering
		\begin{tabular}{| c | l l l l l |} 
			\hline
			 & $S_\text{I}$ & $S_\text{II}$ & $S_\text{III}$ & $S_\text{IV}$ & $S_\text{V}$\\ [0.5ex] 
			\hline
			Lie approach & 0.00413996 & -0.00214115 & 0.000201418 & 0.00015991 & -0.0000117473 \\ 
			\hline
			Kidger \cite{FundOptDesign} & 0.004140 & -0.002141 & 0.000201 & 0.000160 & -0.000012 \\
			\hline
		\end{tabular}
	\caption{Comparison of Seidel sums for the Gaussian doublet.}
	\label{tbl::gaussianDoublet}
	\end{table}

\begin{figure}[htbp!]
		\centering
		\includegraphics[keepaspectratio, width=8cm]{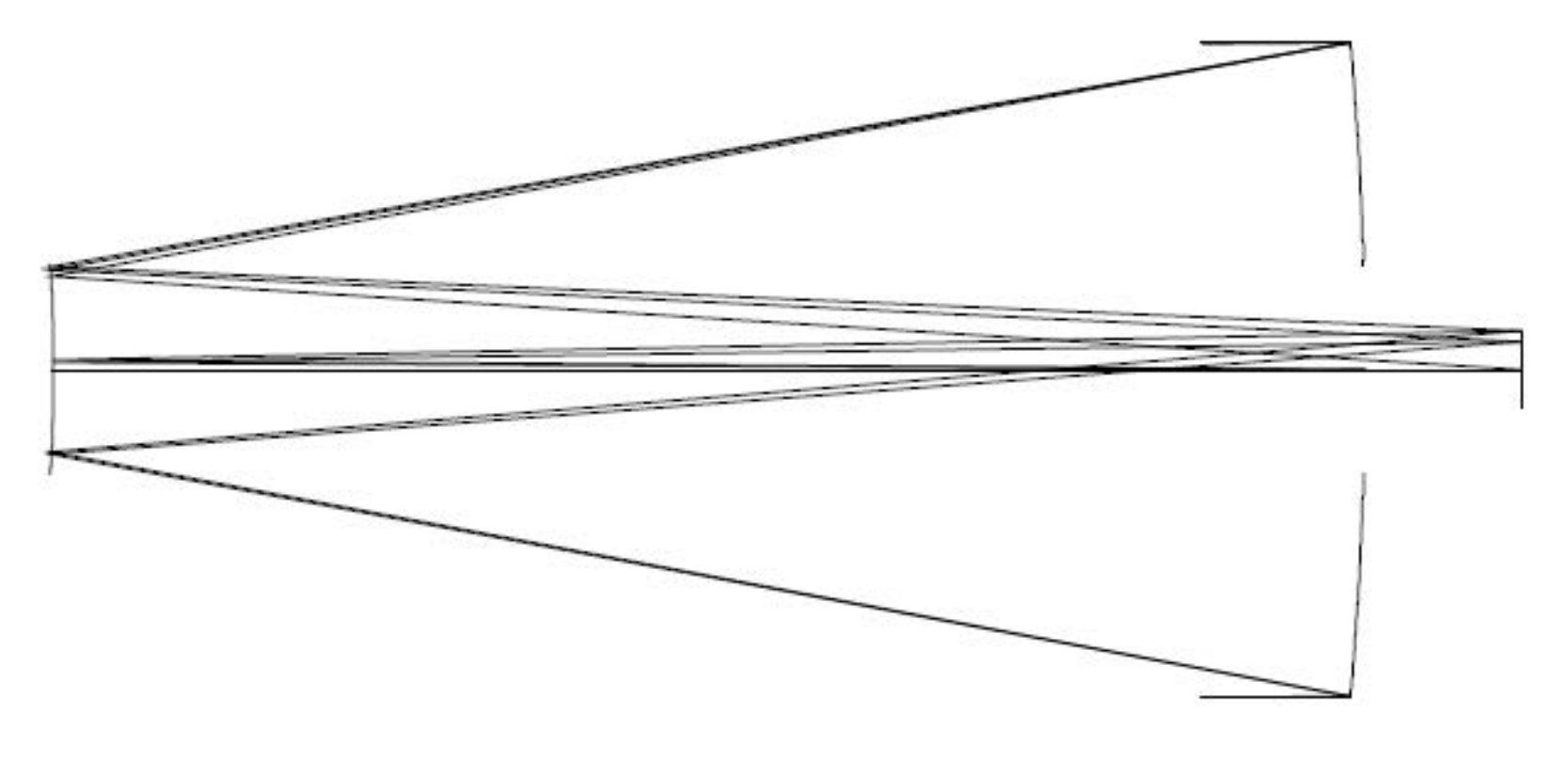}
		\caption{Ritchey-Chr\'etien telescope (Fig. 13.9, \cite{FundOptDesign}).}
		\label{fig::telescope}
	\end{figure}

	The next example is a fully reflective system in the form of a Ritchey-Chr\'etien telescope, see Figure \ref{fig::telescope}. The necessary geometric information is taken from \cite[p. 263]{FundOptDesign}. The procedure to derive the necessary polynomial coefficients is the same as described in Sections \ref{sec::LieForm} and \ref{sec::refrElandSys}. The only change is that the root transformations considered are those for reflection given in Eq.~\eqref{eq::reflRoots}.

	\noindent The Seidel sums are listed in Table \ref{tbl::telescope}.
	\begin{table}[htbp!]
		\centering
		\begin{tabular}{| c | l l l l l |} 
			\hline
			& $S_\text{I}$ & $S_\text{II}$ & $S_\text{III}$ & $S_\text{IV}$ & $S_\text{V}$\\ [0.5ex] 
			\hline
			Lie approach & 0.0000402824 & -0.000164111 & 0.0195427 & 0.0235494 & -0.00206441 \\ 
			\hline
			Kidger \cite{FundOptDesign} &0.000040 & -0.000164 & 0.019537 & 0.023542 & -0.002062 \\
			\hline
		\end{tabular}
		\caption{Comparison of Seidel sums for the Ritchey-Chr\'etien telescope.}
		\label{tbl::telescope}
	\end{table}
	
\begin{figure}[htbp!]
		\centering
		\includegraphics[keepaspectratio, width=8cm]{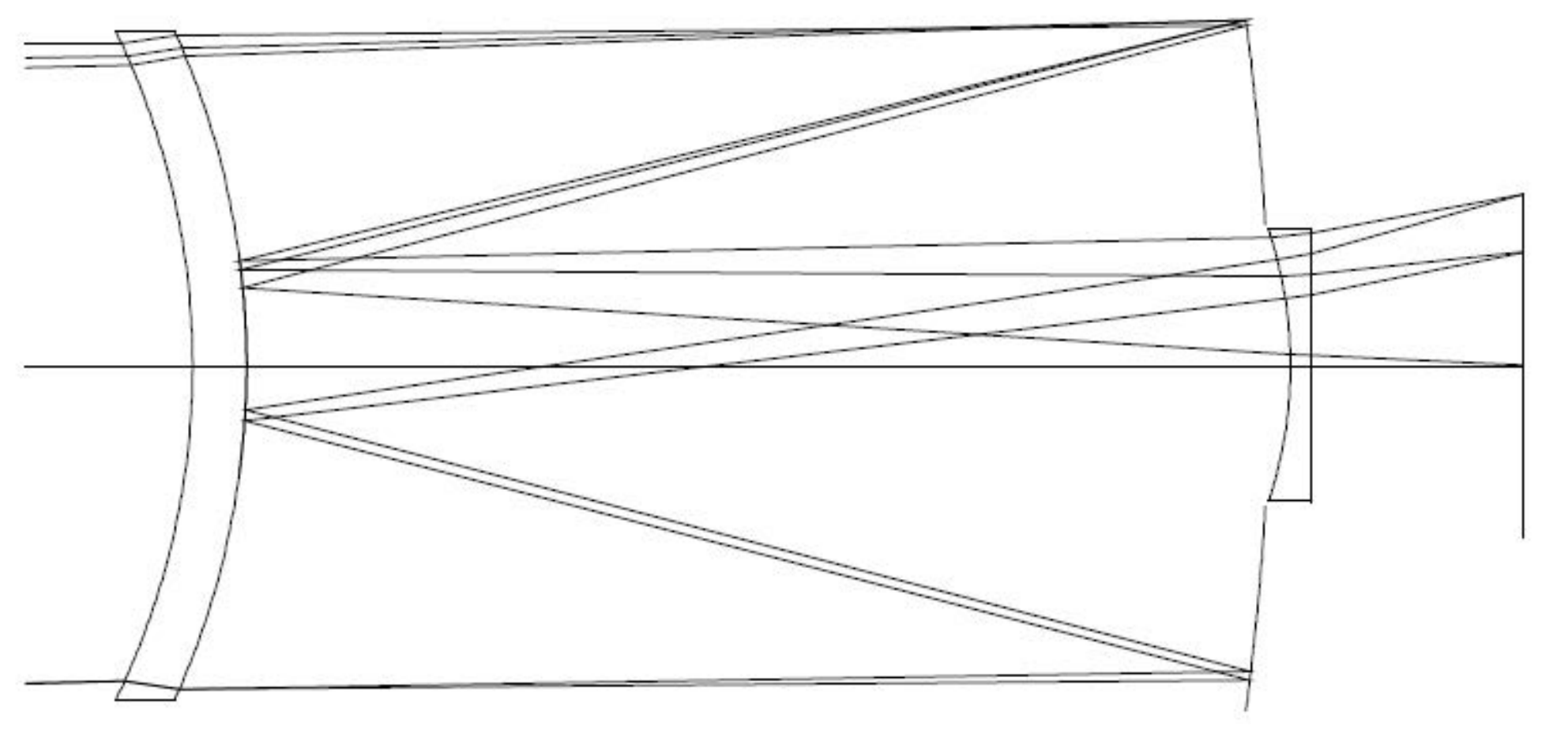}
		\caption{Maksutov-Bouwers Cassegrain system (Fig. 13.20, \cite{FundOptDesign}).}
		\label{fig::catadioptric}
	\end{figure}

	We conclude with a catadioptric, i.e., combined refracting and reflecting, system. The geometric information for the Maksutov-Bouwers Cassegrain optical system, see Figure \ref{fig::catadioptric}, is taken from \cite[p. 272]{FundOptDesign}.
	\noindent The results for the Seidel sums are listed in Table \ref{tbl::catadioptric}.
	\begin{table}[htbp!]
		\centering
		\begin{tabular}{| c | l l l l l |} 
			\hline
			& $S_\text{I}$ & $S_\text{II}$ & $S_\text{III}$ & $S_\text{IV}$ & $S_\text{V}$\\ [0.5ex] 
			\hline
			Lie approach & 0.00500182 & -0.000745778 & -0.000649009 & 0.00286099 & -0.0120072 \\ 
			\hline
			Kidger \cite{FundOptDesign} & 0.005002 & -0.000746 & -0.000649 & 0.002860 & -0.012002 \\
			\hline
		\end{tabular}
		\caption{Comparison of Seidel sums for the  Maksutov-Bouwers Cassegrain system.}
		\label{tbl::catadioptric}
	\end{table}

	We have thus shown how we are able to replicate traditional results often recovering all significant digits.
	
	\section{Conclusions}\label{sec::Conclusions}
	In the presented work, a new way of computing the Seidel coefficients using the Lie algebraic theory has been presented. We briefly introduced the topics of geometrical optics and the Lie algebraic method. After showing how to reformulate the geometrical optics results using Lie algebraic tools, we illustrated how to combine the reformulation of propagation and refraction into arbitrary refractive optical systems with rotational symmetry. The extension to reflective systems can be derived by following the same steps as for refractive systems. The link between the presented Lie framework and the Seidel coefficients has been shown and the property of adding surface contributions is preserved by the Lie algebraic method. The examples given validate our results for refractive, reflective and catadioptric, i.e., combined refractive and reflective, optical systems with rotational symmetry showing good agreement between our and existing results.
	
	The Lie algebraic method offers a mathematical theory and a systematic procedure to link the aberrations of an optical system directly with the geometric parameters of the system itself. All the necessary ray information, i.e., ray heights at interfaces or image height, can be calculated using the Gaussian part of the mapping. The resulting expressions are given explicitly and can be extended to virtually any order of aberrations. The development of a fundamental element makes it straightforward to treat optical systems. In fact, most of the expensive algebraic computations are needed to determine the polynomials associated with the operators. Once these polynomials have been found for the fundamental element, describing an optical system is reduced to concatenating and reordering operators. It is therefore possible to gain additional insight in the interaction of geometries and aberrations of arbitrary order in a closed form, without resorting to ray tracing after each geometric variable change.
	
	The presented methodology also allows to extend these results to off-axis optical systems. The authors intend to apply the Lie algebraic method to reflective systems with only one plane-symmetry, thus allowing for the optical axis to bend. The aim is to derive closed form aberration coefficients for off-axis optical systems. 

\begin{backmatter}
\bmsection{Funding} TKI program ``Photolitho M\&CS" (TKI-HTSM 19.0162)

\bmsection{Acknowledgements} The authors thank Teus Tukker (ASML) for his fruitful remarks.

\bmsection{Disclosures} The authors declare no conflicts of interest.

\bmsection{Data availability} Data underlying the results presented in this paper are available in Ref. \cite{FundOptDesign}.
\end{backmatter}

\bibliography{Bibl}






\end{document}